%% file: atmopaper.tex
\documentclass{elsarticle}
\usepackage{amsmath}
\usepackage{graphicx}
\usepackage{hyperref}
\usepackage{lineno}
\usepackage{multirow}
\usepackage{xspace}

\journal{Astropart. Phys.}

\def \ang      {{\AA}ngstr{\o}m\xspace}
\def \mal      {Malarg\"{u}e\xspace}
\def \pao      {Pierre Auger Observatory\xspace}
\def \xmax     {$X_\text{max}$\xspace}
\def \dxmax    {$\Delta X_\text{max}$\xspace}

\def \relE     {$\Delta E/E$\xspace}
\def \gcmsq    {$\text{g~cm}^{-2}$\xspace}
\def \vaod     {$\tau_{a}(h)$\xspace}
\def \vaodwl   {$\tau_{a}(h,\lambda)$\xspace}

\def \vaodLsr  {$\tau_{a}(h,\lambda_0)$\xspace}
\def \micron   {$\mu\text{m}$\xspace}
\def \trans    {$\mathcal{T}$\xspace}
\def \transa   {$\mathcal{T}_a$\xspace}
\def \transm   {$\mathcal{T}_m$\xspace}

\def \alphaAbs {$\alpha_\text{abs}$\xspace}
\def \nitrogen {N$_2$\xspace}
\def \oxygen   {O$_2$\xspace}
\def \ozone    {O$_3$\xspace}

\begin{document}

\begin{frontmatter}

  \title{A Study of the Effect of \\ 
         Molecular and Aerosol Conditions in the Atmosphere \\
         on Air Fluorescence Measurements at the \\
         \pao}

  \input{author_list_modified.tex}

  \begin{abstract}
    The air fluorescence detector of the \pao is designed to perform
    calorimetric measurements of extensive air showers created by cosmic rays
    of above $10^{18}$~eV.  To correct these measurements for the effects
    introduced by atmospheric fluctuations, the Observatory contains a group of
    monitoring instruments to record atmospheric conditions across the detector
    site, an area exceeding 3,000 km$^2$.  The atmospheric data are used
    extensively in the reconstruction of air showers, and are particularly
    important for the correct determination of shower energies and the depths
    of shower maxima.  This paper contains a summary of the molecular and
    aerosol conditions measured at the \pao since the start of regular
    operations in 2004, and includes a discussion of the impact of these
    measurements on air shower reconstructions.  Between $10^{18}$ and
    $10^{20}$~eV, the systematic uncertainties due to all atmospheric effects
    increase from $4\%$ to $8\%$ in measurements of shower energy, and 
    $4$~\gcmsq to $8$~\gcmsq in measurements of the shower maximum.
  \end{abstract}

\end{frontmatter}
\newpage

\section{Introduction}\label{sec:introduction}

The \pao in \mal, Argentina (69$^{\circ}$ W, 35$^{\circ}$ S, 1400~m a.s.l.) is
a facility for the study of ultra-high energy cosmic rays.  These are primarily
protons and nuclei with energies above $10^{18}$~eV.  Due to the extremely low
flux of high-energy cosmic rays at Earth, the direct detection of such
particles is impractical; but when cosmic rays enter the atmosphere, they
produce extensive air showers of secondary particles.  Using the atmosphere as
the detector volume, the air showers can be recorded and used to reconstruct
the energies, arrival directions, and nuclear mass composition of primary
cosmic ray particles.  However, the constantly changing properties of the
atmosphere pose unique challenges for cosmic ray measurements.

In this paper, we describe the atmospheric monitoring data recorded at the \pao
and their effect on the reconstruction of air showers.  The paper is organized
as follows: Section~\ref{sec:atmocal} contains a review of the observation of
air showers by their ultraviolet light emission, and includes a description of
the \pao and the issues of light production and transmission that arise when
using the atmosphere to make cosmic ray measurements.  The specifics of light
attenuation by aerosols and molecules are described in
Section~\ref{sec:prod_and_trans}.  An overview of local molecular measurements
is given in Section~\ref{sec:molecular_effects}, and in
Section~\ref{sec:measurements} we discuss cloud-free aerosol measurements
performed at the Observatory.  The impact of these atmospheric measurements on
the reconstruction of air showers is explored in
Section~\ref{sec:aerosol_effects}.  Cloud measurements with infrared cameras
and backscatter lidars are briefly described in Section~\ref{sec:future}.
Conclusions are given in Section~\ref{sec:conclusion}.

\input atmocal

\input atmotrans

\input measmol

\input measaer

\input hybridres

\input futuredevel

\section{Conclusions}\label{sec:conclusion}

  A large collection of atmospheric monitors is operated at the \pao to provide
  frequent observations of molecular and aerosol conditions across the
  detector.  These data are used to estimate light scattering losses between
  air showers and the FD telescopes, to correct air shower light production for
  various weather effects, and to prevent cloud-obscured data from distorting
  estimates of the shower energies, shower maxima, and the detector aperture.

  In this paper, we have described the various light production and
  transmission effects due to molecules and aerosols.  These effects have been
  converted into uncertainties in the hybrid reconstruction.  Most of the
  reported uncertainties are systematic, not only due to the use of local
  empirical models to describe the atmosphere --- such as the monthly molecular
  profiles --- but also because of the nature of the atmospheric uncertainties
  --- such as the systematics-dominated and highly correlated aerosol optical
  depth profiles.

  Molecular measurements are vital for the proper determination of light
  production in air showers, and molecular scattering is the dominant term in
  the description of atmospheric light propagation.  However, the time
  variations in molecular scattering conditions are small relative to
  variations in the aerosol component.  The inherent variability in aerosol
  conditions can have a significant impact on the data if aerosol measurements
  are not incorporated into the reconstruction.  Because the highest energy air
  showers are viewed at low elevation angles and through long distances in the
  aerosol boundary layer, aerosol effects become increasingly important at high
  energies.
  
  Efforts are currently underway to reduce the systematic uncertainties due to
  the atmosphere, with particularly close attention paid to the uncertainties
  in energy and \xmax.  The shoot-the-shower program will improve the time
  resolution of atmospheric measurements, and increase the identification of
  atmospheric inhomogeneities that can affect observations of showers with the
  FD telescopes.

\section{Acknowledgments}\label{sec:acknowledgments}
\input acknowledgments

\bibliography{atmopaper}
\bibliographystyle{elsarticle-num}

\end{document}

%% file: author_list_modified.tex
\author{\noindent{\bf The Pierre Auger Collaboration} \\
J.~Abraham$^{8}$, 
P.~Abreu$^{71}$, 
M.~Aglietta$^{54}$, 
C.~Aguirre$^{12}$, 
E.J.~Ahn$^{87}$, 
D.~Allard$^{31}$, 
I.~Allekotte$^{1}$, 
J.~Allen$^{90}$, 
J.~Alvarez-Mu\~{n}iz$^{78}$, 
M.~Ambrosio$^{48}$
% ...
\\
\noindent$^{1}$ Centro At\'{o}mico Bariloche and Instituto Balseiro (CNEA-
UNCuyo-CONICET), San Carlos de Bariloche, Argentina \\
% ...
$^{8}$ National Technological University, Faculty Mendoza 
(CONICET/CNEA), Mendoza, Argentina \\
% ...
$^{12}$ Universidad Catolica de Bolivia, La Paz, Bolivia \\
% ...
$^{31}$ Laboratoire AstroParticule et Cosmologie (APC), 
Universit\'{e} Paris 7, CNRS-IN2P3, Paris, France \\
% ...
$^{48}$ Universit\`{a} di Napoli ``Federico II'' and Sezione INFN, 
Napoli, Italy \\
% ...
$^{54}$ Istituto di Fisica dello Spazio Interplanetario (INAF),
 Universit\`{a} di Torino and Sezione INFN, Torino, Italy \\
% ...
$^{71}$ LIP and Instituto Superior T\'{e}cnico, Lisboa, Portugal \\
% ...
$^{78}$ Universidad de Santiago de Compostela, Spain \\
% ...
$^{87}$ Fermilab, Batavia, IL, USA \\
% ...
$^{90}$ New York University, New York, NY, USA \\
% ...
\noindent (\ddag) Deceased \\
(a) at Konan University, Kobe, Japan \\
(b) On leave of absence at the Instituto Nacional de Astrofisica, Optica y Electronica \\
(c) at Caltech, Pasadena, USA \\
(d) at Hawaii Pacific University \\
}

%% file: atmocal.tex
\section{Cosmic Ray Observations using Atmospheric Calorimetry}\label{sec:atmocal}

  \subsection{The Air Fluorescence Technique}
 
  The charged secondary particles in extensive air showers produce copious
  amounts of ultraviolet light -- of order $10^{10}$ photons per meter near the
  peak of a $10^{19}$~eV shower.  Some of this light is due to nitrogen
  fluorescence, in which molecular nitrogen excited by a passing shower emits
  photons isotropically into several dozen spectral bands between $300$ and
  $420$~nm.  A much larger fraction of the shower light is emitted as Cherenkov
  photons, which are strongly beamed along the shower axis.  With square-meter
  scale telescopes and sensitive photodetectors, the UV emission from the
  highest energy air showers can be observed at distances in excess of $30$~km
  from the shower axis.

  The flux of fluorescence photons from a given point on an air shower track is
  proportional to $dE/dX$, the energy loss of the shower per unit slant depth
  $X$ of traversed atmosphere~\cite{Bunner:1967,Arqueros:2008cx}. The emitted
  light can be used to make a calorimetric estimate of the energy of the
  primary cosmic ray~\cite{Baltrusaitis:1985mx,Unger:2008uq}, after a small
  correction for the ``missing energy'' not contained in the electromagnetic
  component of the shower.  Note that a large fraction of the light received
  from a shower may be contaminated by Cherenkov photons.  However, if the
  Cherenkov fraction is carefully estimated, it can also be used to measure the
  longitudinal development of a shower~\cite{Unger:2008uq}.

  The fluorescence technique can also be used to determine cosmic ray
  composition.  The slant depth at which the energy deposition rate, $dE/dX$,
  reaches its maximum value, denoted \xmax, is correlated with the mass of the
  primary particle~\cite{Linsley:1977,McComb:1982}.  Showers generated by light
  nuclei will, on average, penetrate more deeply into the atmosphere than
  showers initiated by heavy particles of the same energy, although the exact
  behavior is dependent on details of hadronic interactions and must be
  inferred from Monte Carlo simulations.  By observing the UV light from air
  showers, it is possible to estimate the energies of individual cosmic rays,
  as well as the average mass of a cosmic ray data set.
  
  \subsection{Challenges of Atmospheric Calorimetry}

  The atmosphere is responsible for producing light from air showers.  Its
  properties are also important for the transmission efficiency of light from
  the shower to the air fluorescence detector.  The atmosphere is variable, and
  so measurements performed with the air fluorescence technique must be
  corrected for changing conditions, which affect both light production and
  transmission.
  
  For example, extensive balloon measurements conducted at the
  \pao~\cite{Keilhauer:2005ja} and a study using radiosonde data from various
  geographic locations~\cite{BADC:website} have shown that the altitude profile
  of the atmospheric depth, $X(h)$, typically varies by $\sim5$~\gcmsq from one
  night to the next.  In extreme cases, the depth can change by $20$~\gcmsq on
  successive nights, which is similar to the differences in depth between the
  seasons~\cite{Wilczynska:2006wt}.  The largest variations are comparable to
  the \xmax resolution of the Auger air fluorescence detector, and could
  introduce significant biases into the determination of \xmax if not properly
  measured.  Moreover, changes in the bulk properties of the atmosphere such as
  air pressure $p$, temperature $T$, and humidity $u$ can have a significant
  effect on the rate of nitrogen fluorescence emission~\cite{Keilhauer:2005nk},
  as well as light transmission.

  In the lowest $15$ km of the atmosphere where air shower measurements occur,
  sub-$\mu$m to mm-sized aerosols also play an important role in modifying the
  light transmission.  Most aerosols are concentrated in a boundary layer that
  extends about $1$~km above the ground, and throughout most of the
  troposphere, the ultraviolet extinction due to aerosols is typically several
  times smaller than the extinction due to
  molecules~\cite{Tegen:1996,Baltensperger:2003,Kinne:2006}.  However, the
  variations in aerosol conditions have a greater effect on air shower
  measurements than variations in $p$, $T$, and $u$, and during nights with
  significant haze, the light flux from distant showers can be reduced by
  factors of $3$ or more due to aerosol attenuation.  The vertical density
  profile of aerosols, as well as their size, shape, and composition, vary
  quite strongly with location and in time, and depending on local particle
  sources (dust, smoke, etc.) and sinks (wind and rain), the density of
  aerosols can change substantially from hour to hour.  If not properly
  measured, such dynamic conditions can bias shower reconstructions.
  
  \subsection{The Pierre Auger Observatory}

  The Pierre Auger Observatory contains two cosmic ray detectors.  The first is
  a Surface Detector (SD) comprising 1600 water Cherenkov stations to observe
  air shower particles that reach the ground~\cite{Allekotte:2007sf}.  The
  stations are arranged on a triangular grid of $1.5$~km spacing, and the full
  SD covers an area of 3,000~km$^2$.  The SD has a duty cycle of nearly 100\%,
  allowing it to accumulate high-energy statistics at a much higher rate than
  was possible at previous observatories.
  
  Operating in concert with the SD is a Fluorescence Detector (FD) of 24 UV
  telescopes~\cite{fdpaper}.  The telescopes are arranged to overlook the SD
  from four buildings around the edge of the ground array.  Each of the four FD
  buildings contains six telescopes,  and the total field of view at each site
  is $180^\circ$ in azimuth and $1.8^\circ-29.4^\circ$ in elevation.  The main
  component of a telescope is a spherical mirror of area $11$~m$^2$ that
  directs collected light onto a camera of 440 hexagonal photomultipliers
  (PMTs).  One photomultiplier ``pixel'' views approximately
  $1.5^\circ\times1.5^\circ$ of the sky, and its output is digitized at
  $10$~MHz.  Hence, every PMT camera can record the development of air showers
  with $100$~ns time resolution.
  
  The FD is only operated during dark and clear conditions, when the shower UV
  signal is not overwhelmed by moonlight or blocked by low clouds or rain.
  These limitations restrict the FD duty cycle to $\sim10\%-15$\%, but unlike
  the SD, the FD data provide calorimetric estimates of shower energies.
  Simultaneous SD and FD measurements of air showers, known as hybrid
  observations, are used to calibrate the absolute energy scale of the SD,
  reducing the need to calibrate the SD with shower simulations.  The hybrid
  operation also dramatically improves the geometrical and longitudinal profile
  reconstruction of showers measured by the FD, compared to showers observed by
  the FD
  alone~\cite{Sommers:1995dm,AbuZayyad:2000ay,Mostafa:2006id,Dawson:2007di}.
  This high-quality hybrid data set is used for all physics analyses based on
  the FD.
  
  \begin{figure}[ht]
    \begin{center}
      \includegraphics*[width=0.6\textwidth,clip]{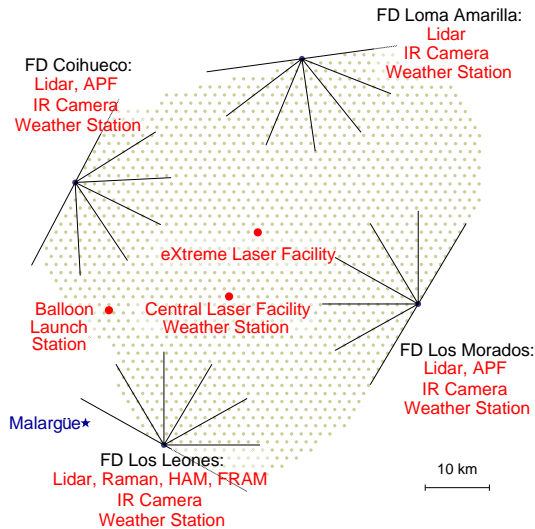}
      \caption{\label{fig:auger_detector} \slshape The Surface Detector
               stations and Fluorescence Detector sites of the \pao.  Also
               shown are the locations of \mal and the atmospheric monitoring
               instruments operating at the Observatory (see text for details).}
    \end{center}
  \end{figure}

  To remove the effect of atmospheric fluctuations that would otherwise impact
  FD measurements, an extensive atmospheric monitoring program is carried out
  at the \pao.  A list of monitors and their locations relative to the FD
  buildings and SD array are shown in fig.~\ref{fig:auger_detector}.
  Atmospheric conditions at ground level are measured by a network of weather
  stations at each FD site and in the center of the SD; these provide updates
  on ground-level conditions every five minutes.  In addition, regular
  meteorological radiosonde flights (one or two per week) are used to measure
  the altitude profiles of atmospheric pressure, temperature, and other bulk
  properties of the air.  The weather station monitoring and radiosonde flights
  are performed day or night, independent of the FD data acquisition.

  During the dark periods suitable for FD data-taking, hourly measurements
  of aerosols are made using the FD telescopes, which record vertical UV laser
  tracks produced by a Central Laser Facility (CLF) deployed on site since
  2003~\cite{Fick:2006yn}.  These measurements are augmented by data from
  lidar stations located near each FD building~\cite{BenZvi:2006xb}, a
  Raman lidar at one FD site, and the eXtreme Laser Facility (or XLF, named for
  its remote location) deployed in November 2008.  Two Aerosol Phase Function
  Monitors (APFs) are used to determine the aerosol scattering properties of
  the atmosphere using collimated horizontal light beams produced by Xenon
  flashers~\cite{BenZvi:2007px}.  Two optical telescopes --- the Horizontal
  Attenuation Monitor (HAM) and the (F/ph)otometric Robotic Telescope for
  Atmospheric Monitoring (FRAM) --- record data used to determine the
  wavelength dependence of the aerosol
  attenuation~\cite{BenZvi:2007it,BenZvi:2007uj}.  Finally, clouds are measured
  hourly by the lidar stations, and infrared cameras on the roof of
  each FD building are used to record the cloud coverage in the FD field of
  view every five minutes~\cite{Valore:2009}.

%% file: atmotrans.tex
\section{The Production of Light by the Shower and its Transmission through the
Atmosphere}\label{sec:prod_and_trans}

  Atmospheric conditions impact on both the production and transmission of UV
  shower light recorded by the FD.  The physical conditions of the molecular
  atmosphere have several effects on fluorescence light production, which we
  summarize in Section~\ref{subsec:weath_light_prod}.  We treat light
  transmission, outlined in Section~\ref{subsec:weath_trans}, primarily as a
  single-scattering process characterized by the atmospheric optical depth
  (Sections~\ref{subsubsec:mol_optical_depth} and
  \ref{subsubsec:aero_optical_depth}) and scattering angular dependence
  (Section~\ref{subsubsec:mol_aero_scattering}).  Multiple scattering
  corrections to atmospheric transmission are discussed in
  Section~\ref{subsubsec:mult_scat}.

  \subsection{The Effect of Weather on Light Production}\label{subsec:weath_light_prod}

  The yields of light from the Cherenkov and fluorescence emission processes
  depend on the physical conditions of the gaseous mixture of molecules in the
  atmosphere.  The production of Cherenkov light is the simpler of the two
  cases, since the number of photons emitted per charged particle per meter per
  wavelength interval depends only on the refractive index of the atmosphere
  $n(\lambda,p,T)$.  The dependence of this quantity on pressure, temperature,
  and wavelength $\lambda$ can be estimated analytically, and so the effect of
  weather on the light yield from the Cherenkov process are relatively simple
  to incorporate into air shower reconstructions.

  The case of fluorescence light is more complex, not only because it is
  necessary to consider additional weather effects on the light yield, but also
  due to the fact that several of these effects can be determined only by
  difficult experimental measurements (see
  \cite{Kakimoto:1995pr,Nagano:2004am,Abbasi:2008zz,Bohacova:2008vg} and
  references in \cite{Arqueros:2008en}).

  One well-known effect of the weather on light production is the collisional
  quenching of fluorescence emission, in which the radiative transitions of
  excited nitrogen molecules are suppressed by molecular collisions.  The rate
  of collisions depends on pressure and temperature, and the form of this
  dependence can be predicted by kinetic gas
  theory~\cite{Bunner:1967,Nagano:2004am}.  However, the cross section for
  collisions is itself a function of temperature, which introduces an
  additional term into the $p$ and $T$ dependence of the yield.   The
  temperature dependence of the cross section cannot be predicted \textsl{a
  priori}, and must be determined with laboratory
  measurements~\cite{Ave:2007xh}.

  Water vapor in the atmosphere also contributes to collisional quenching, and
  so the fluorescence yield has an additional dependence on the absolute
  humidity of the atmosphere.  This dependence must also be determined
  experimentally, and its use as a correction in shower reconstructions using
  the fluorescence technique requires regular measurements of the altitude
  profile of humidity.  A full discussion of these effects is beyond the scope
  of this paper, but detailed descriptions are available
  in~\cite{Arqueros:2008cx,Keilhauer:2005nk,Keilhauer:2008sy}.  We will
  summarize the estimates of their effect on shower energy and \xmax in
  Section~\ref{subsec:mol_uncertainties}.

  \subsection{The Effect of Weather on Light Transmission}\label{subsec:weath_trans}

  The attenuation of light along a path through the atmosphere between a light
  source and an observer can be expressed as a transmission coefficient \trans,
  which gives the fraction of light not absorbed or scattered along the path.
  If the optical thickness (or optical depth) of the path is $\tau$, then
  \trans is estimated using the Beer-Lambert-Bouguer law:
  \begin{linenomath}
    \begin{equation}\label{eq:trans_od}
      \mathcal{T}=e^{-\tau}.
    \end{equation}
  \end{linenomath}

  The optical depth of the air is affected by the density and composition of
  molecules and aerosols, and can be treated as the sum of molecular and
  aerosol components: $\tau=\tau_m+\tau_a$.  The optical depth is a function of
  wavelength and the orientation of a path within the atmosphere.  However, if
  the atmospheric region of interest is composed of horizontally uniform
  layers, then the full spatial dependence of $\tau$ reduces to an altitude
  dependence, such that $\tau\equiv\tau(h,\lambda)$.  For a slant path elevated
  at an angle $\varphi$ above the horizon, the light transmission along the
  path between the ground and height $h$ is
  \begin{linenomath}
  \begin{equation}\label{eq:aerotrans}
    \mathcal{T}(h,\lambda,\varphi) = 
      e^{-\tau(h,\lambda)/\sin{\varphi}}.
  \end{equation}
  \end{linenomath}

  In an air fluorescence detector, a telescope recording isotropic 
  fluorescence emission of intensity $I_0$ from a source of light along a
  shower track will observe an intensity
  \begin{linenomath}
    \begin{equation}\label{eq:intensity}
      I=I_0\cdot
      \mathcal{T}_m\cdot\mathcal{T}_a\cdot(1+H.O.)\cdot\frac{\Delta\Omega}{4\pi}\text{,}
    \end{equation}
  \end{linenomath}
  where $\Delta\Omega$ is the solid angle subtended by the telescope diaphragm
  as seen from the light source.  The molecular and aerosol transmission factor
  \transm$\cdot$\transa primarily represents single-scattering of photons out
  of the field of view of the telescope.  In the ultraviolet range used for air
  fluorescence measurements, the absorption of light is much less important
  than scattering~\cite{Tegen:1996,Seinfeld:2006}, although there are some
  exceptions discussed in Section~\ref{subsubsec:mol_optical_depth}.  The term
  $H.O.$ is a higher-order correction to the Beer-Lambert-Bouguer law that
  accounts for the single and multiple scattering of Cherenkov and fluorescence
  photons into the field of view.

  To estimate the transmission factors and scattering corrections needed in
  eq.~\eqref{eq:intensity}, it is necessary to measure the vertical height
  profile and wavelength dependence of the optical depth $\tau(h,\lambda)$, as
  well as the angular distribution of light scattered from atmospheric
  particles, also known as the phase function $P(\theta)$.  For these
  quantities, the contributions due to molecules and aerosols are considered
  separately.

  \subsubsection{The Optical Depth of Molecules}\label{subsubsec:mol_optical_depth}

    The probability per unit length that a photon will be scattered or absorbed
    as it moves through the atmosphere is given by the total volume extinction
    coefficient
    \begin{linenomath}
    \begin{equation}\label{eq:ext_coeff}
      \alpha_\text{ext}(h,\lambda) = \alpha_\text{abs}(h,\lambda) 
      + \beta(h,\lambda)\text{,}
    \end{equation}
    \end{linenomath}
    where \alphaAbs and $\beta$ are the coefficients of absorption and
    scattering, respectively.  The vertical optical depth between a telescope
    at ground level and altitude $h$ is the integral of the atmospheric
    extinction along the path:
    \begin{linenomath}
    \begin{equation}\label{eq:optical_depth_definition}
      \tau(h,\lambda)=\int_{h_\text{gnd}}^h\alpha_\text{ext}(h',\lambda)dh'.
    \end{equation}
    \end{linenomath}

    Molecular extinction in the near UV is primarily an elastic scattering
    process, since the Rayleigh scattering of light by molecular nitrogen
    (\nitrogen) and oxygen (\oxygen) dominates inelastic scattering and
    absorption~\cite{Killinger:1987}.  For example, the Raman scattering cross
    sections of \nitrogen and \oxygen are approximately $10^{-30}$~cm$^{-2}$
    between $300-420$~nm \cite{Burris:1992}, much smaller than the Rayleigh
    scattering cross section of air ($\sim10^{-27}$~cm$^{-2}$) at these
    wavelengths~\cite{Bucholtz:1995}.  Moreover, while \oxygen is an important
    absorber in the deep UV, its absorption cross section is effectively zero
    for wavelengths above $240$~nm~\cite{Seinfeld:2006}.  Ozone (\ozone)
    molecules absorb light in the UV and visible bands, but \ozone is mainly
    concentrated in a high-altitude layer above the atmospheric volume used for
    air fluorescence measurements~\cite{Seinfeld:2006}.
    
    Therefore, for the purpose of air fluorescence detection, the total
    molecular extinction $\alpha_\text{ext}^m(h,\lambda)$ simply reduces to the
    scattering coefficient $\beta_m(h,\lambda)$.  At standard temperature and
    pressure, molecular scattering can be defined analytically in terms of the
    Rayleigh scattering cross section \cite{Bucholtz:1995,Naus:2000}:
    \begin{linenomath}
    \begin{equation}\label{eq:molecular_scattering_coeff}
      \beta^\text{STP}_m(h,\lambda)\equiv\beta_s(\lambda) =N_s\sigma_\text{R}(\lambda)
        = \frac{24\pi^3}{N_s\lambda^4}
          \left(\frac{n_s^2(\lambda)-1}{n_s^2(\lambda)+2}\right)^2
          \frac{6+3\rho(\lambda)}{6-7\rho(\lambda)}.
    \end{equation}
    \end{linenomath}
    In this expression, $N_s$ is the molecular number density under standard
    conditions and $n_s(\lambda)$ is the index of refraction of air.  The
    depolarization ratio of air, $\rho(\lambda)$, is determined by the
    asymmetry of \nitrogen and \oxygen molecules, and its value is
    approximately $0.03$ in the near UV~\cite{Bucholtz:1995}.  The wavelength
    dependence of these quantities means that between $300$~nm and $420$~nm,
    the wavelength dependence of molecular scattering shifts from the classical
    $\lambda^{-4}$ behavior to an effective value of $\lambda^{-4.2}$.

    Since the atmosphere is an ideal gas, the altitude dependence of the
    scattering coefficient can be expressed in terms of the vertical
    temperature and pressure profiles $T(h)$ and $p(h)$,
    \begin{linenomath}
    \begin{equation}\label{eq:scatter_profile}
      \alpha_\text{ext}^m(h,\lambda)\equiv\beta_m(h,\lambda) = \beta_s(\lambda)
        \frac{p(h)}{p_s} \frac{T_s}{T(h)}\text{,}
    \end{equation}
    \end{linenomath}
    where $T_s$ and $p_s$ are standard temperature and
    pressure~\cite{Bucholtz:1995}.  Given the profiles $T(h)$ and $p(h)$ 
    obtained from balloon measurements or local climate models, the vertical
    molecular optical depth is estimated via numerical integration of
    equations~\eqref{eq:optical_depth_definition} and
    \eqref{eq:scatter_profile}.

  \subsubsection{The Optical Depth of Aerosols}\label{subsubsec:aero_optical_depth}

    The picture is more complex for aerosols than for molecules because in
    general it is not possible to calculate the total aerosol extinction
    coefficient analytically.  The particulate scattering theory of Mie, for
    example, depends on the simplifying assumption of spherical
    scatterers~\cite{Mie:1908}, a condition which often does not hold in the
    field\footnote{Note that in spite of this, aerosol scattering is often
    referred to as ``Mie scattering.''}. Moreover, aerosol scattering depends
    on particle composition, which can change quite rapidly depending on the
    wind and weather conditions.  

    Therefore, knowledge of the aerosol transmission factor \transa depends on
    frequent field measurements of the vertical aerosol optical depth \vaodwl.
    Like other aerosol properties, the altitude profile of \vaodwl can change
    dramatically during the course of a night.  However, in general \vaodwl
    increases rapidly with $h$ only in the first few kilometers above ground
    level, due to the presence of mixed aerosols in the planetary boundary
    layer.
  
\begin{figure}[ht]
      \begin{center}
        \includegraphics*[width=\textwidth,clip]{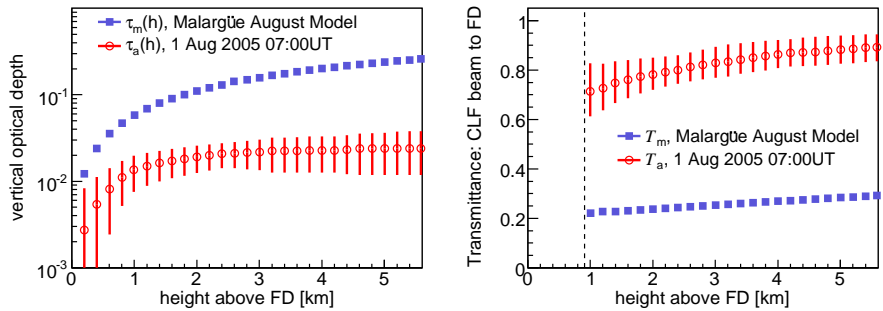}
        \caption{\label{fig:vaod_transmission} \slshape 
                 Left: a vertical aerosol optical depth profile
                 $\tau_a(h,355~\text{nm})$ measured using the FD at Los Leones
                 with vertical laser shots from the CLF ($26$~km distance).
                 The uncertainties are dominated by systematic effects and are
                 highly correlated.  Also shown is the monthly average
                 molecular optical depth $\tau_m(h,355~\text{nm})$.  Right:
                 molecular and aerosol light transmission factors for the
                 atmosphere between the vertical CLF laser beam and the Los
                 Leones FD.  The dashed line at $1$~km indicates the lower 
                 edge of the FD field of view at this distance (see
                 Section~\ref{subsubsec:central_laser_facility} for details).}
      \end{center}
    \end{figure}

    In the lower atmosphere, the majority of aerosols are concentrated in the
    mixing layer.  The thickness of the mixing layer is measured from the
    prevailing ground level in the region, and its height roughly follows the
    local terrain (excluding small hills and escarpments).  This gives the
    altitude profile of \vaodwl a characteristic shape: a nearly linear
    increase at the lowest heights, followed by a flattening as the aerosol
    density rapidly decreases with altitude.
    Figure~\ref{fig:vaod_transmission} depicts an optical depth profile
    inferred using vertical laser shots from the CLF at $355$~nm viewed from
    the FD site at Los Leones.  The profile, corresponding to a moderately
    clear atmosphere, can be considered typical of this location.  Also shown
    is the aerosol transmission coefficient between points along the vertical
    laser beam and the viewing FD, corresponding to a ground distance of
    $26$~km.  
  
    The wavelength dependence of \vaodwl depends on the wavelength of the
    incident light and the size of the scattering aerosols.  A conventional
    parameterization for the dependence is a power law due to
    \ang~\cite{Angstrom:1929},
    \begin{linenomath}
    \begin{equation}\label{eq:wavelength_dependence}
      \tau_a(h,\lambda) = 
        \tau(h,\lambda_0)\cdot\left(\frac{\lambda_0}{\lambda}\right)^\gamma,
    \end{equation}
    \end{linenomath}
    where $\gamma$ is known as the \ang exponent.  The exponent is also
    measured in the field, and the measurements are normalized to the value of
    the optical depth at a reference wavelength $\lambda_0$.  The normalization
    point used at the Auger Observatory is the wavelength of the Central Laser
    Facility, $\lambda_0=355$~nm, approximately in the center of the nitrogen
    fluorescence spectrum.
    
    The \ang exponent is determined by the size distribution of scattering
    aerosols, such that smaller particles have a larger exponent --- eventually
    reaching the molecular limit of $\gamma\approx4$ --- while larger particles
    give rise to a smaller $\gamma$ and thus a more ``wavelength-neutral''
    attenuation~\cite{McCartney:1976,Schuster:2006}.  For example, in a review
    of the literature by Eck et al.~\cite{Eck:1999}, aerosols emitted from
    burning vegetation and urban and industrial areas are observed to have a
    relatively large \ang coefficient ($\gamma=1.41\pm0.35$).  These
    environments are dominated by fine ($<1~\mu$m) ``accumulation mode''
    particles, or aerodynamically stable aerosols that do not coalesce or
    settle out of the atmosphere.  In desert environments, where coarse
    ($>1~\mu$m) particles dominate, the wavelength dependence is almost
    negligible~\cite{Eck:1999,Whittet:1987}.

  \subsubsection{Angular Dependence of Molecular and Aerosol Scattering}\label{subsubsec:mol_aero_scattering}

    Only a small fraction of the photons emitted from an air shower arrive at a
    fluorescence detector without scattering.  The amount of scattering must be
    estimated during the reconstruction of the shower, and so the scattering
    properties of the atmosphere need to be well understood.
    
    For both molecules and aerosols, the angular dependence of scattering is
    described by normalized angular scattering cross sections, which give the
    probability per unit solid angle $P(\theta)=\sigma^{-1}d\sigma/d\Omega$
    that light will scatter out of the beam path through an angle $\theta$.
    Following the convention of the atmospheric literature, this work will
    refer to the normalized cross sections as the molecular and aerosol phase
    functions.
    
    The molecular phase function $P_m(\theta)$ can be estimated analytically,
    with its key feature being the symmetry in the forward and backward
    directions.  It is proportional to the $(1+\cos^2{\theta})$ factor of the
    Rayleigh scattering theory, but in air there is a small correction factor
    $\delta\approx1\%$ due to the anisotropy of the N$_2$ and O$_2$ molecules
    \cite{Bucholtz:1995}:
    \begin{linenomath}
    \begin{equation}\label{eq:mol_phase_function}
      P_m(\theta) = \frac{3}{16\pi(1+2\delta)}
      \left(1+3\delta + (1-\delta)\cos^2{\theta}\right).
    \end{equation}
    \end{linenomath}
    
    The aerosol phase function $P_a(\theta)$, much like the aerosol optical
    depth, does not have a general analytical solution, and in fact its
    behavior as a function of $\theta$ is quite complex.  Therefore, one is
    often limited to characterizing the gross features of the light scattering
    probability distribution, which is sufficient for the purposes of air
    fluorescence detection.  In general, the angular distribution of light
    scattered by aerosols is very strongly peaked in the forward direction,
    reaches a minimum near $90^{\circ}$, and has a small backscattering
    component.  It is reasonably approximated by the
    parameterization~\cite{BenZvi:2007px,Fishburne:1976,Riewe:1978}
    \begin{linenomath}
    \begin{equation}\label{eq:aero_phase_function}
      P_a(\theta) = \frac{1-g^2}{4\pi}\cdot
                  \left(\frac{1}{(1+g^2-2g\cos{\theta})^{3/2}}+
                  f\frac{3\cos^2{\theta}-1}{2(1+g^2)^{3/2}}
                  \right).
    \end{equation}
    \end{linenomath}
    The first term, a Henyey-Greenstein scattering function~\cite{Henyey:1941},
    corresponds to forward scattering; and the second term --- a second-order
    Legendre polynomial, chosen so that it does not affect the normalization of
    $P_a(\theta)$ --- accounts for the peak at large $\theta$ typically found
    in the angular distribution of aerosol-scattered light.  The quantity
    $g=\langle\cos{\theta}\rangle$ measures the asymmetry of scattering, and
    $f$ determines the relative strength of the forward and backward scattering
    peaks.  The parameters $f$ and $g$ are observable quantities which depend
    on local aerosol characteristics.  

  \subsubsection{Corrections for Multiple Scattering}\label{subsubsec:mult_scat}

  As light propagates from a shower to the FD, molecular and aerosol scattering
  can remove photons that would otherwise travel along a direct path toward an
  FD telescope.  Likewise, some photons with initial paths outside the detector
  field of view can be scattered back into the telescope, increasing the
  apparent intensity and angular width of the shower track.

  During the reconstruction of air showers, it is convenient to consider the
  addition and subtraction of scattered photons to the total light flux in
  separate stages.  The subtraction of light is accounted for in the
  transmission coefficients \transm and \transa of eq.~\eqref{eq:intensity}.
  Given the shower geometry and measurements of atmospheric scattering
  conditions, the estimation of \transm and \transa is relatively
  straightforward.  However, the addition of light due to atmospheric
  scattering is less simple to calculate, due to the contributions of multiple
  scattering.  Multiple scattering has no universal analytical description, and
  those analytical solutions which do exist are only valid under restrictive
  assumptions that do not apply to typical FD viewing
  conditions~\cite{Roberts:2005xv}.

  A large fraction of the flux of photons from air showers recorded by an FD
  telescope can come from multiply-scattered light, particularly within the
  first few kilometers above ground level, where the density of scatterers is
  highest.  In poor viewing conditions, $10\%-15\%$ of the photons arriving
  from the lower portion of a shower track may be due to multiple scattering.
  Since these contributions cannot be neglected, a number of Monte Carlo
  studies have been carried out to quantify the multiply-scattered component of
  recorded shower signals under realistic atmospheric
  conditions~\cite{Roberts:2005xv,Pekala:2003wu,Giller:2005,Pekala:2009fe}.
  The various simulations indicate that multiple scattering grows with optical
  depth and distance from the shower.  Based on these results,
  Roberts~\cite{Roberts:2005xv} and Pekala et al.~\cite{Pekala:2009fe} have
  developed parameterizations of the fraction of multiply-scattered photons in
  the shower image.  Both parameterizations are implemented in the FD event
  reconstruction, and their effect on estimates of the shower energy and shower
  maximum are described in section~\ref{subsec:ms_correction}.

%% file: measmol.tex
\section{Molecular Measurements at the \pao}\label{sec:molecular_effects}

  \subsection{Profile Measurements with Weather Stations and Radiosondes}\label{subsec:profile_measurements}

  The vertical profiles of atmospheric parameters (pressure, temperature, etc.)
  vary with geographic location and with time so that a global static model of
  the atmosphere is not appropriate for precise shower studies.  At a given
  location, the daily variation of the atmospheric profiles can be as large as
  the variation in the seasonal average conditions.  Therefore, daily
  measurements of atmospheric profiles are desirable.

  \begin{figure}[ht]
    \begin{center}
      \includegraphics*[width=0.85\textwidth,clip]{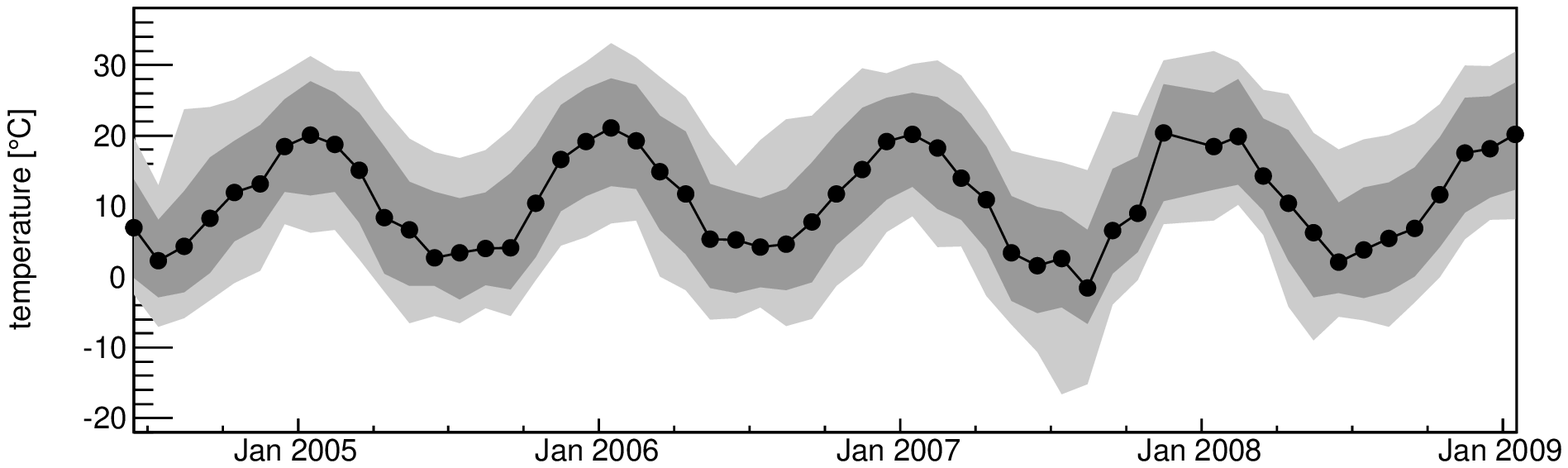}
      \includegraphics*[width=0.85\textwidth,clip]{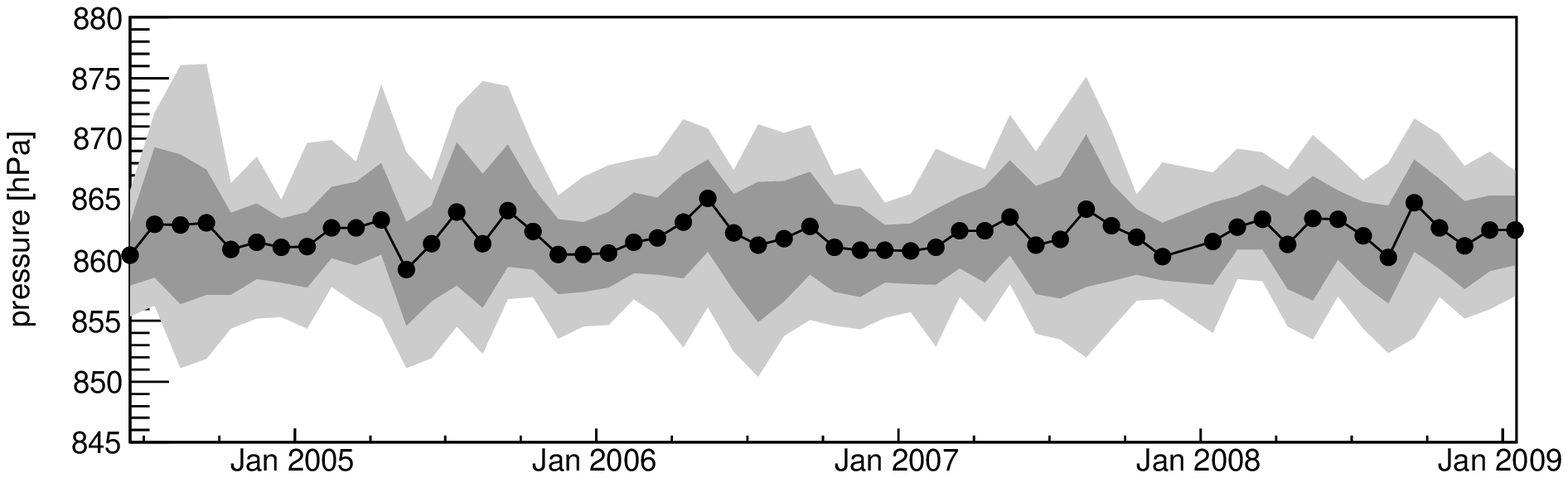}
      \includegraphics*[width=0.85\textwidth,clip]{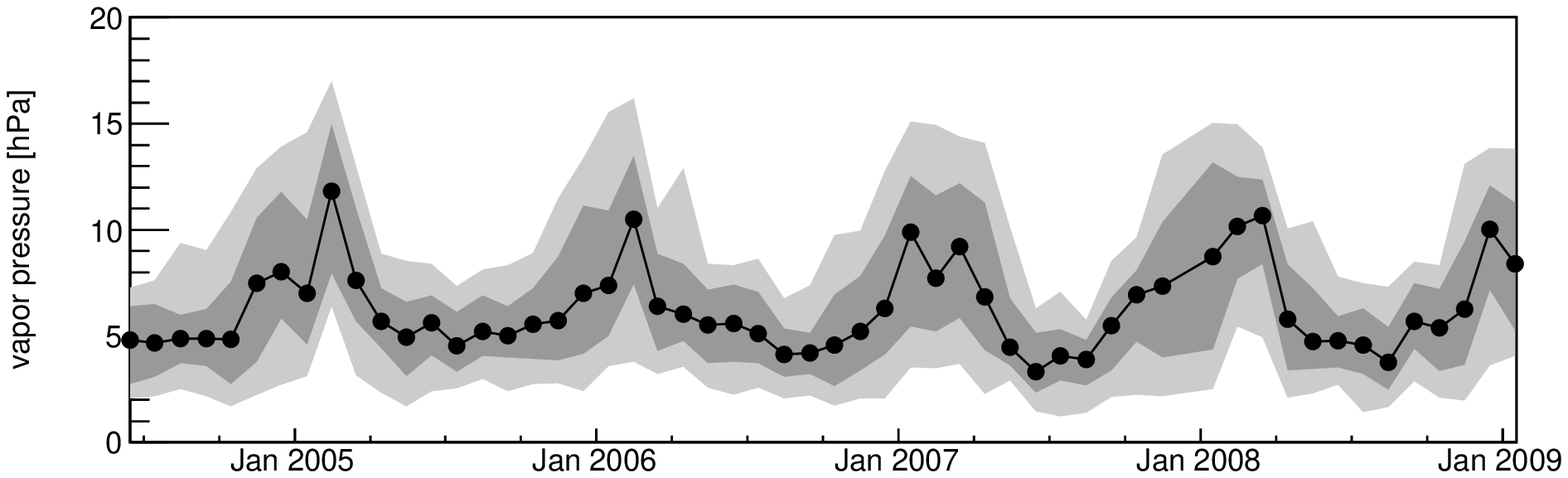}
      \caption{\label{fig:weather_ground} \slshape 
               Monthly median ground temperature, pressure, and water vapor
               pressure observed at the CLF weather station ($1.4$~km above
               sea level), showing the distributions of 68\% and 95\% of the
               measurements as dark and light gray contours, respectively.  The
               vapor pressure has been calculated using measurements of
               the temperature and relative humidity.}
    \end{center}
  \end{figure}

  Several measurements of the molecular component of the atmosphere are
  performed at the \pao.  Near each FD site and the CLF, ground-based weather
  stations are used to record the temperature, pressure, relative humidity, and
  wind speed every five minutes.  The first weather station was commissioned at
  Los Leones in January 2002, followed by stations at the CLF (June 2004), Los
  Morados (May 2007), and Loma Amarilla (November 2007).  The station at
  Coihueco is installed but not currently operational.  Data from the CLF
  station are shown in fig.~\ref{fig:weather_ground}; the measurements are
  accurate to $0.2-0.5^\circ$C in temperature, $0.2-0.5$~hPa in pressure, and
  $2\%$ in relative humidity~\cite{Campbell:website}.  The pressure and
  temperature data from the weather stations are used to monitor the weather
  dependence of the shower signal observed by the
  SD~\cite{Bleve:2007mb,Bleve:2009}.  They can also be used to characterize the
  horizontal uniformity of the molecular atmosphere, which is assumed in
  eq.~\eqref{eq:aerotrans}.
  
  Of more direct interest to the FD reconstruction are measurements of the
  altitude dependence of the pressure and temperature, which can be used in
  eq.~\eqref{eq:scatter_profile} to estimate the vertical molecular optical
  depth.  These measurements are performed with balloon-borne radiosonde
  flights, which began in mid-2002 and are currently launched one or two times
  per week.  The radiosonde measurements include relative humidity and wind
  data recorded about every $20$~m up to an average altitude of $25$~km, well
  above the fiducial volume of the fluorescence detectors.  The accuracy of the
  measurements are approximately $0.2^\circ\text{C}$ for temperature,
  $0.5-1.0$~hPa for pressure, and $5\%$ for relative
  humidity~\cite{GRAW:website}.
  
  \begin{figure}[ht]
    \begin{center}
      \includegraphics*[width=\textwidth,clip]{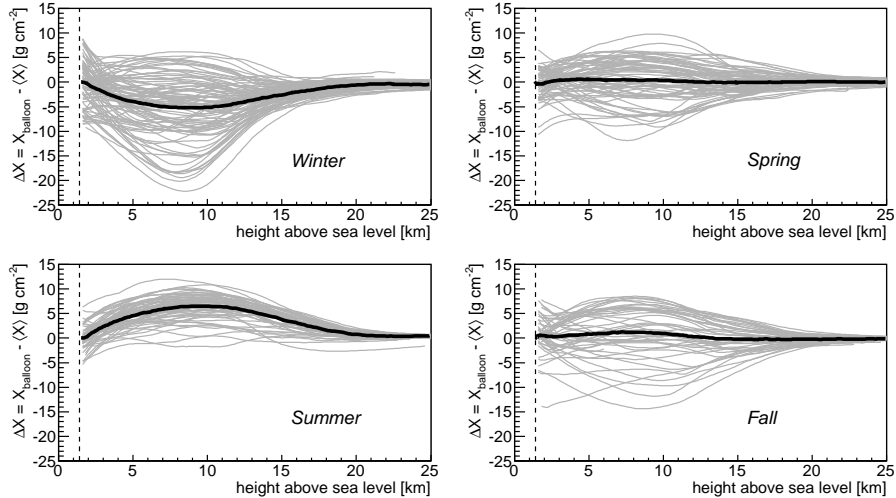}
      \caption{\label{fig:balloonX} \slshape 
               Radiosonde measurements of the depth profile above \mal recorded
               during 261 balloon flights between 2002 and 2009.  The data are
               plotted as deviations from the average profile of all 261
               flights, and are grouped by season.  The dark lines indicate the
               seasonal averages, and the vertical dashed lines correspond to
               the height of \mal above sea level.}
    \end{center}
  \end{figure}

  The balloon observations demonstrate that daily variations in the temperature
  and pressure profiles depend strongly on the season, with more stable
  conditions during the austral summer than in winter~\cite{Keilhauer:2005ja}.
  The atmospheric depth profile $X(h)$ exhibits significant altitude-dependent
  fluctuations.  The largest daily fluctuations are typically $5$~\gcmsq
  observed at ground level, increasing to $10-15$~\gcmsq between $6$ and
  $12$~km altitude.  The seasonal differences between summer and winter can be
  as large as $20$~\gcmsq on the ground, increasing to $30$~\gcmsq at higher
  altitudes (fig.~\ref{fig:balloonX}).

  \subsection{Monthly Average Models}\label{subsec:monthly_models}

  Balloon-borne radiosondes have proven to be a reliable means of measuring the
  state variables of the atmosphere, but nightly balloon launches are too
  difficult and expensive to carry out with regularity in \mal.  Therefore, it
  is necessary to sacrifice some time resolution in the vertical profile
  measurements and use models which quantify the average molecular profile over
  limited time intervals.
  
  \begin{figure}[ht]
    \begin{center}
      \includegraphics*[width=\textwidth,clip]{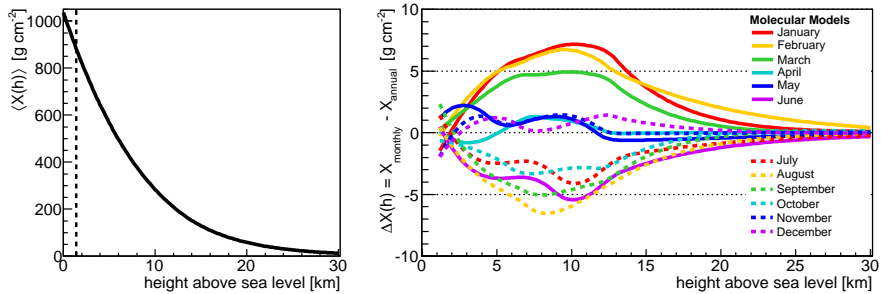}
      \caption{\label{fig:monthly_profiles} \slshape 
               Left: average profile $X(h)$ above \mal, with the altitude of the
               site indicated by the vertical dotted line.  Right:
               deviation of the monthly mean values of $X(h)$ from the yearly
               average as a function of month.  Data are from the mean monthly
               weather models (updated through 2009).}
    \end{center}
  \end{figure}

  Such time-averaged models have been generated for the FD reconstruction using
  $261$ local radiosonde measurements conducted between August 2002 and
  December 2008.  The monthly profiles include average values for the
  atmospheric depth, density, pressure, temperature, and humidity as a function
  of altitude.  Figure~\ref{fig:monthly_profiles} depicts a plot of the annual
  mean depth profile $X(h)$ in \mal, as well as the deviation of the monthly
  model profiles from the annual average.  The uncertainties in the monthly
  models, not shown in the figure, represent the typical range of conditions
  observed during the course of each month.  At ground level, the RMS
  uncertainties are approximately $3$~\gcmsq in austral summer and $6$~\gcmsq
  during austral winter; near $10$~km altitude, the uncertainties are
  $4$~\gcmsq in austral summer and $8$~\gcmsq in austral winter.
  
  The use of monthly averages rather than daily measurements introduces
  uncertainties into measurements of shower energies $E$ and shower maxima
  \xmax; the magnitudes of the effects are estimated in
  Section~\ref{subsec:mol_uncertainties}.
  
  \subsection{Horizontal Uniformity of the Molecular Atmosphere}\label{subsec:horizontal_molecular}

  The assumption of horizontally uniform atmospheric layers implied by
  equation~\eqref{eq:aerotrans} reduces the estimate of atmospheric
  transmission to a simple geometrical calculation, but the deviation of the
  atmosphere from true horizontal uniformity introduces some systematic error
  into the transmission. An estimate of this deviation is required to
  calculate its impact on air shower reconstruction.
  
  \begin{figure}[ht]
    \begin{center}
      \includegraphics*[width=0.85\textwidth,clip]{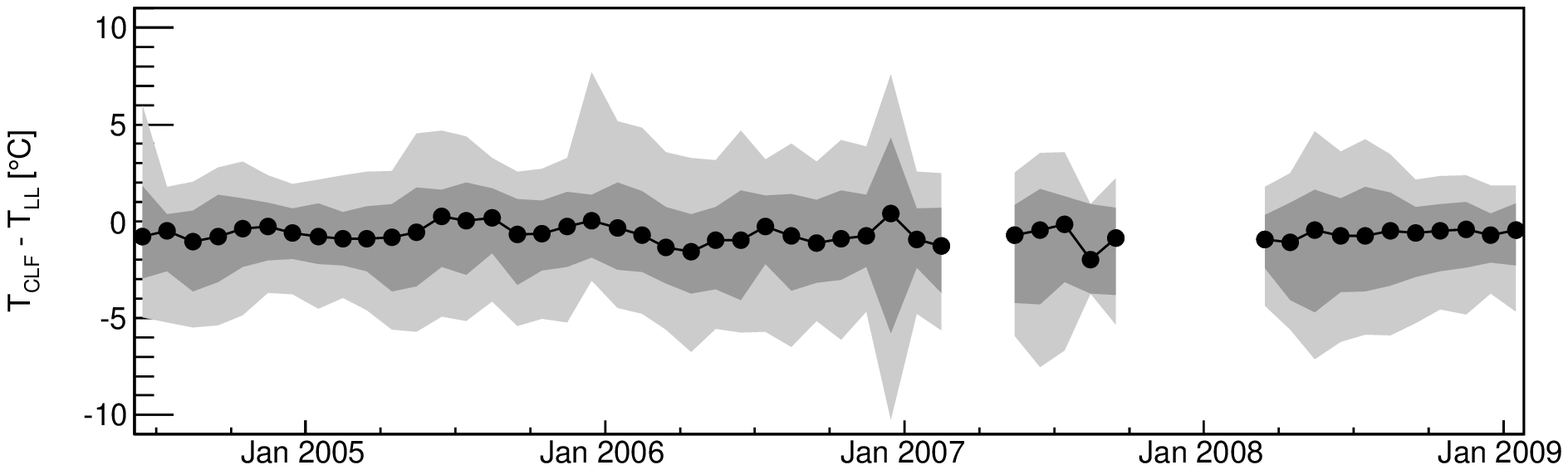}
      \includegraphics*[width=0.85\textwidth,clip]{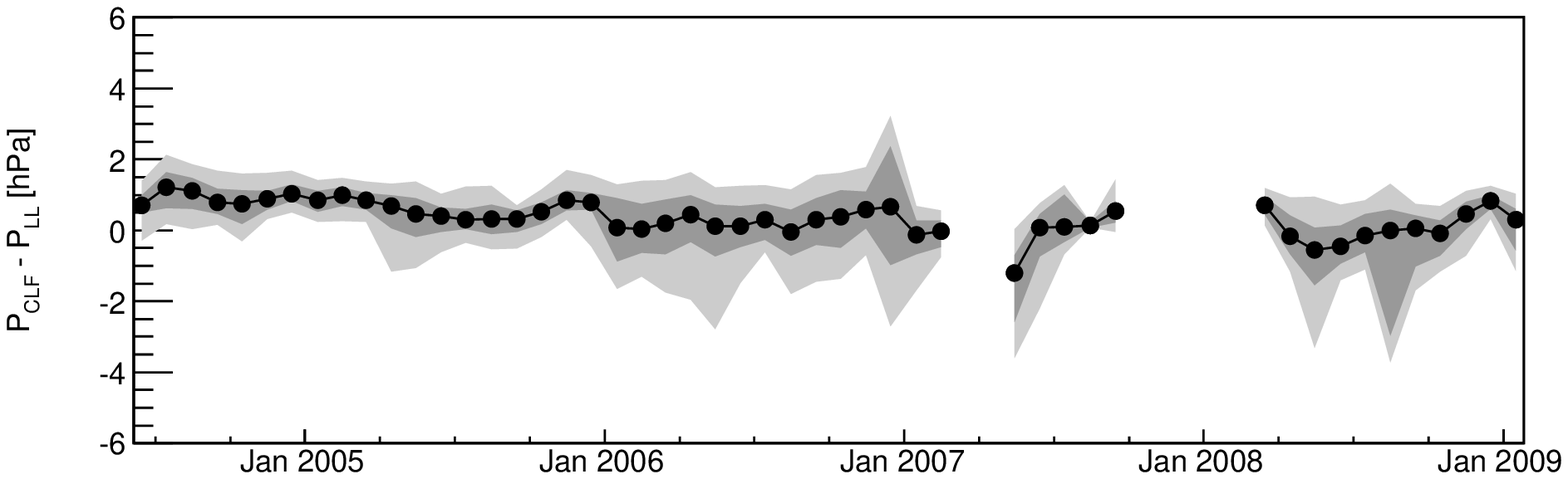}
      \includegraphics*[width=0.85\textwidth,clip]{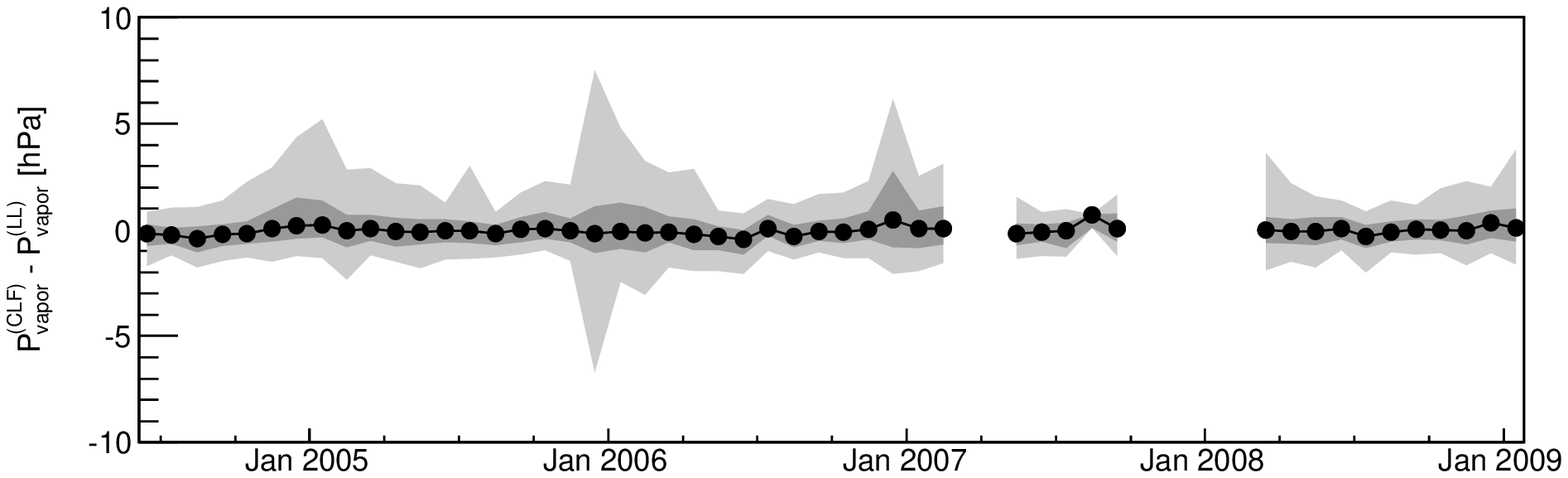}
      \caption{\label{fig:diff_weather_ground} \slshape 
        Monthly differences in the ground temperature, pressure, and vapor
        pressure observed with the weather stations at Los Leones (LL) and the
        CLF.  The dark and light gray contours contain 68\% and 95\% of the
        measurement differences.  Gaps in the comparison during 2007 were
        caused by equipment failures in the station at Los Leones.}
    \end{center}
  \end{figure}

  For the molecular component of the atmosphere, the data from different
  ground-based weather stations provide a convenient, though limited, check of
  weather differences across the Observatory.  For example, the differences
  between the temperature, pressure, and vapor pressure measured using the
  weather stations at Los Leones and the CLF are plotted in
  fig.~\ref{fig:diff_weather_ground}.  The altitude difference between the
  stations is approximately $10$~m, and they are separated by $26$~km, or
  roughly half the diameter of the SD.  Despite the large horizontal separation
  of the sites, the measurements are in close agreement.  Note that the
  differences in the vapor pressure are larger than the differences in total
  pressure, due to the lower accuracy of the relative humidity measurements.
  
  It is quite difficult to check the molecular uniformity at higher altitudes,
  with, for example, multiple simultaneous balloon launches.  The measurements
  from the network of weather stations at the Observatory are currently the
  only indications of the long-term uniformity of molecular conditions across
  the site.  Based on these observations, the molecular atmosphere is treated
  as uniform.

%% file: measaer.tex
\section{Aerosol Measurements at the \pao}\label{sec:measurements}

  Several instruments are deployed at the \pao to observe aerosol scattering
  properties.  The aerosol optical depth is estimated using UV laser
  measurements from the CLF, XLF, and scanning lidars
  (Section~\ref{subsec:aod_measurements}); the aerosol phase function is
  determined with APF monitors (Section~\ref{subsec:aero_scat_measurements});
  and the wavelength dependence of the aerosol optical depth is measured with
  data recorded by the HAM and FRAM telescopes
  (Section~\ref{subsec:aero_wl_measurements}).

  \subsection{Optical Depth Measurements}\label{subsec:aod_measurements}

    \subsubsection{The Central Laser Facility}\label{subsubsec:central_laser_facility}

\begin{figure}[ht]
      \begin{center}
        \includegraphics*[width=0.4\textwidth,clip]{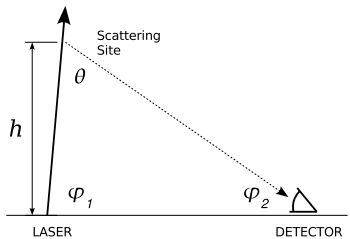}
        \caption{\label{fig:laser_geometry} \slshape CLF laser and FD geometry.
          Vertical shots ($\varphi_1=90^\circ$) are used for the measurement of
          \vaodLsr, with $\lambda_0=355$~nm.}
      \end{center}
    \end{figure}

    The CLF produces calibrated laser ``test beams'' from its location in the
    center of the Auger surface detector~\cite{Fick:2006yn,Wiencke:2006hg}.
    Located between $26$ and $39$~km from the FD telescopes, the CLF contains a
    pulsed $355$~nm laser that fires a depolarized beam in an quarter-hourly
    sequence of vertical and inclined shots.  Light is scattered out of the
    laser beam, and a small fraction of the scattered light is collected by the
    FD telescopes.  With a nominal energy of $7$~mJ per pulse, the light
    produced is roughly equal to the amount of fluorescence light generated by
    a $10^{20}$~eV shower.  The CLF-FD geometry is shown in
    fig.~\ref{fig:laser_geometry}.

    The CLF has been in operation since late 2003.  Every quarter-hour during
    FD data acquisition, the laser fires a set of 50 vertical shots.  The
    relative energy of each vertical shot is measured by two ``pick-off''
    energy probes, and the light profiles recorded by the FD telescopes are
    normalized by the probe measurements to account for shot-by-shot changes in
    the laser energy.  The normalized profiles are then averaged to obtain
    hourly light flux profiles, in units of photons~m$^{-2}$~mJ$^{-1}$ per
    100~ns at the FD entrance aperture~\cite{Fick:2006yn}.  The hourly profiles
    are determined for each FD site, reflecting the fact that aerosol
    conditions may not be horizontally uniform across the Observatory during
    each measurement period.

    \begin{figure}[ht]
      \begin{center}
      \includegraphics*[width=0.85\textwidth,clip]{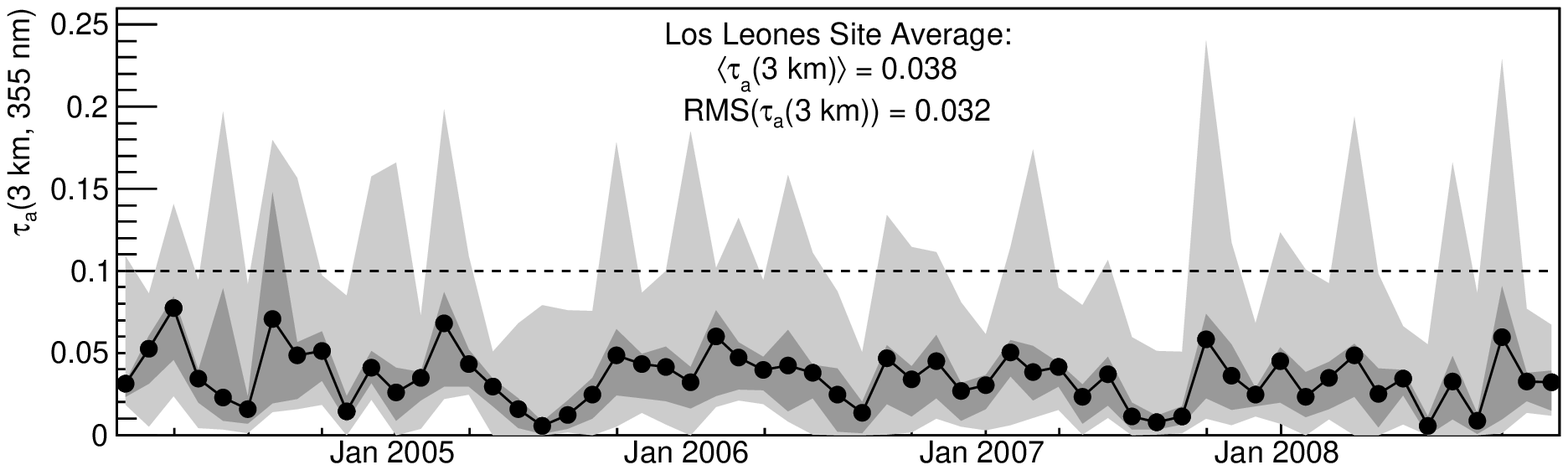}
      \includegraphics*[width=0.85\textwidth,clip]{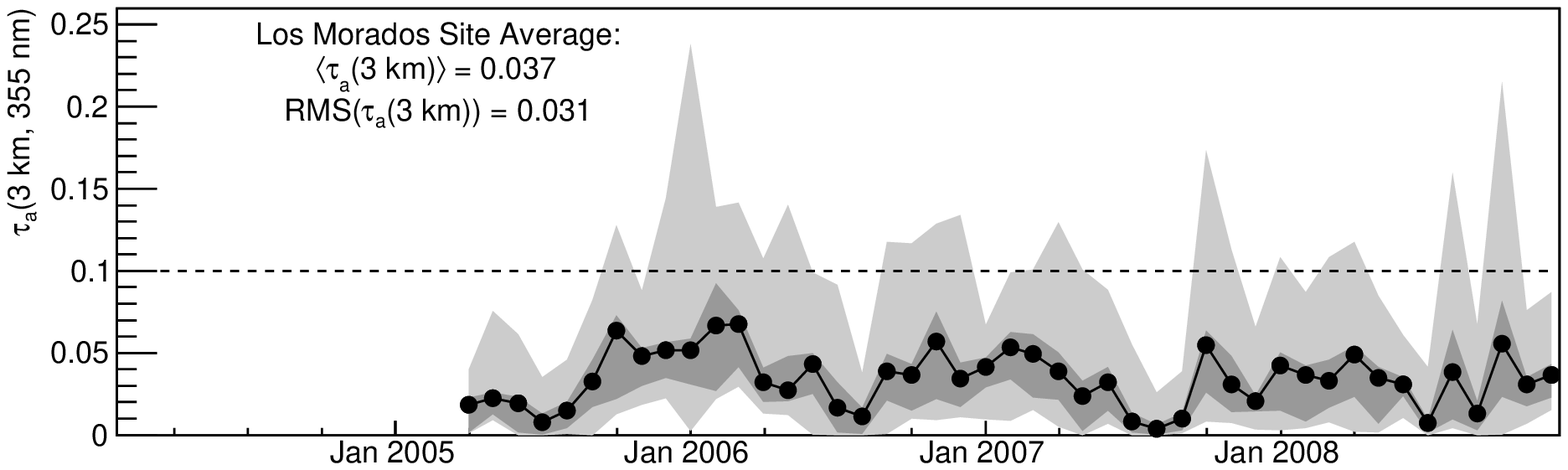}
      \includegraphics*[width=0.85\textwidth,clip]{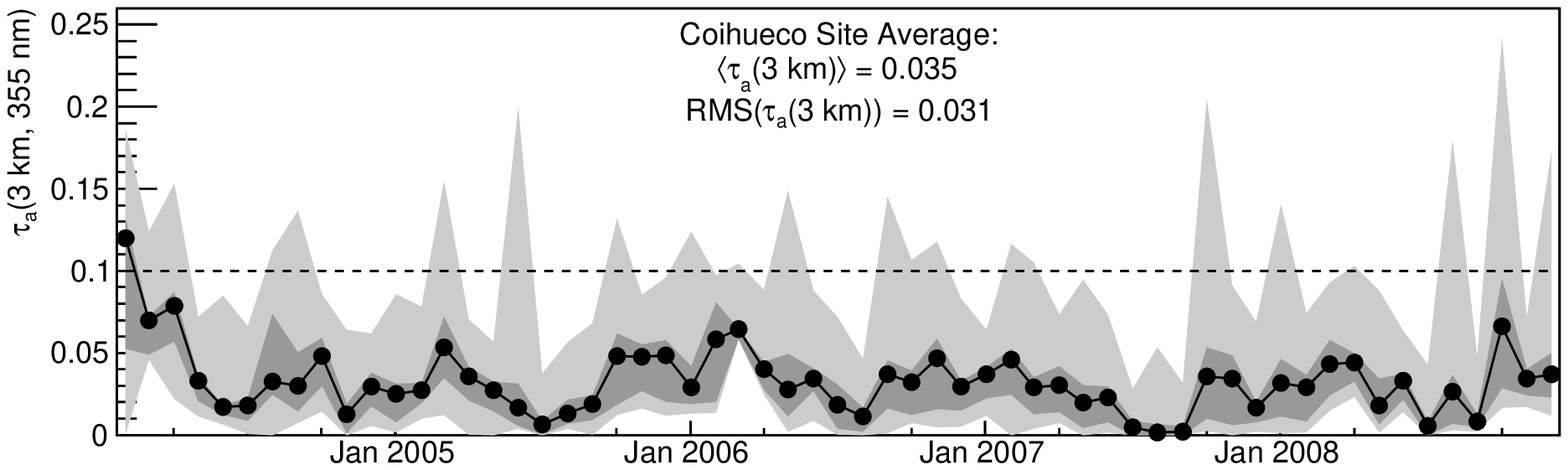}
      \caption{\label{fig:clfVaod3km} \slshape
               Monthly median CLF measurements of the aerosol optical depth
               3~km above the fluorescence telescopes at Los Leones,
               Los Morados, and Coihueco (January 2004 -- December 2008).
               Measurements from Loma Amarilla are not currently available.
               The dark and light contours contain 68\% and 95\% of the
               measurements, respectively.  Hours with optical depths above 0.1
               (dashed lines) are characterized by strong haze, and are cut
               from the FD analysis.}
      \end{center}
    \end{figure}

    It is possible to determine the vertical aerosol optical depth \vaodLsr
    between the CLF and an FD site by normalizing the observed light flux with
    a ``molecular reference'' light profile.  The molecular references are
    simply averaged CLF laser profiles that are observed by the FD telescopes
    during extremely clear viewing conditions with negligible aerosol
    attenuation.  The references can be identified by the fact that the laser
    light flux measured by the telescopes during clear nights is larger than
    the flux on nights with aerosol attenuation (after correction for the
    relative calibration of the telescopes).  Clear-night candidates can also
    be identified by comparing the shape of the recorded light profile against
    a laser simulation using only Rayleigh scattering~\cite{Valore:2009}.  The
    candidate nights are then validated by measurements from the APF monitors
    and lidar stations.
    
    A minimum of three consecutive clear hours are used to construct each
    reference profile.  Once an hourly profile is normalized by a
    clear-condition reference, the attenuation of the remaining light is due
    primarily to aerosol scattering along the path from the CLF beam to the
    telescopes.  The optical depth \vaodLsr can be extracted from the
    normalized hourly profiles using the methods described
    in~\cite{Abbasi:2005gi}.

    Note that the lower elevation limit of the FD telescopes ($1.8^\circ$)
    means that the lowest $1$~km of the vertical laser beam is not within
    the telescope field of view (see fig.~\ref{fig:vaod_transmission}).  While
    the CLF can be used to determine the total optical depth between the ground
    and $1$~km, the vertical distribution of aerosols in the lowest part of the
    atmosphere cannot be observed.  Therefore, the optical depth in this region
    is constructed using a linear interpolation between ground level, where
    $\tau_a$ is zero, and $\tau_a(1~\text{km},\lambda_0)$.

    The normalizations used in the determination of \vaodLsr mean that the
    analysis does not depend on the absolute photometric calibration of either
    the CLF or the FD, but instead on the accuracy of relative calibrations of
    the laser and the FD telescopes.

    The sources of uncertainty that contribute to the normalized hourly
    profiles include the clear night references ($3\%$)\footnote{The value
    $3\%$ contains the statistical and calibration uncertainties in a given
    reference profile, but does not describe an uncertainty in the selection of
    the reference.  This uncertainty will be quantified in a future end-to-end
    analysis of CLF data using simulated laser shots.}, uncertainties in the FD
    relative calibration ($3\%$), and the accuracy of the laser energy
    measurement ($3\%$).  Statistical fluctuations in the hourly average light
    profiles contribute additional relative uncertainties of $1\%-3\%$ to the
    normalized hourly light flux.  The uncertainties in \vaodLsr plotted in
    fig.~\ref{fig:vaod_transmission} derive from these sources, and are highly
    correlated due to the systematic uncertainties.
    
    Between January 2004 and December 2008, over 6,000 site-hours of optical
    depth profiles have been analyzed using measurements of more than one
    million CLF shots.  Figure~\ref{fig:clfVaod3km} depicts the distribution of
    \vaod recorded using the FD telescopes at Los Leones, Los Morados, and
    Coihueco.  The data $3$~km above ground level are shown, since this
    altitude is typically above the aerosol mixing layer.  A moderate seasonal
    dependence is apparent in the aerosol distributions, with austral summer
    marked by more haze than winter.  The distributions are asymmetric, with
    long tails extending from the relatively clear conditions 
    ($\tau_{a}(3~\text{km})<0.04$) characteristic of most hours to periods of
    significant haze ($\tau_{a}(3~\text{km})>0.1$). 
    
    Approximately $5\%$ of CLF measurements have optical depths greater than
    0.1.  To avoid making very large corrections to the expected light flux
    from distant showers, these hours are typically not used in the FD
    analysis.

    \subsubsection{Lidar Observations}

    In addition to the CLF, four scanning lidar stations are operated
    at the \pao to record \vaodLsr from every FD site~\cite{BenZvi:2006xb}.
    Each station has a steerable frame that holds a pulsed $351$~nm laser,
    three parabolic mirrors, and three PMTs.  The frame is mounted atop a
    shipping container which contains data acquisition electronics.  The
    station at Los Leones includes a separate, vertically-pointing Raman lidar
    test system, which can be used to detect aerosols and the relative
    concentration of N$_2$ and O$_2$ in the atmosphere.

    During FD data acquisition, the lidar telescopes sweep the sky in a set
    hourly pattern, pulsing the laser at 333~Hz and observing the backscattered
    light with the optical receivers.  By treating the altitude distribution of
    aerosols near each lidar station as horizontally uniform, \vaodLsr can be
    estimated from the differences in the backscattered laser signal recorded
    at different zenith angles~\cite{Filipcic:2002ba}.  When non-uniformities
    such as clouds enter the lidar sweep region, the optical depth can still be
    determined up to the altitude of the non-uniformity.
    
    Since the lidar hardware and measurement techniques are independent of the
    CLF, the two systems have essentially uncorrelated systematic
    uncertainties.  With the exception of a short hourly burst of horizontal
    shots toward the CLF and a shoot-the-shower mode (Section~\ref{subsec:sts})
    ~\cite{BenZvi:2006xb}, the lidar sweeps occur outside the FD field of view
    to avoid triggering the detector with backscattered laser light.  Thus, for
    many lidar sweeps, the extent to which the lidars and CLF measure similar
    aerosol profiles depends on the true horizontal uniformity of aerosol
    conditions at the Observatory.

    \begin{figure}
      \begin{center}
      \includegraphics*[width=\textwidth,clip]{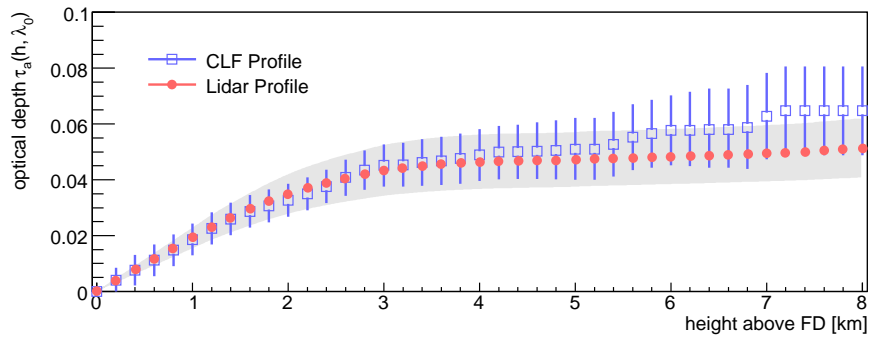}
      \caption{\label{fig:lidar_clf_profile_comparison} \slshape
        An hourly aerosol optical depth profile observed by the CLF and
        the Coihueco lidar station for relatively dirty conditions in
        December 2006.  The gray band depicts the systematic uncertainty
        in the lidar aerosol profile.}
      \end{center}
    \end{figure}

    Figure~\ref{fig:lidar_clf_profile_comparison} shows a lidar measurement of
    \vaodLsr with vertical shots and the corresponding CLF aerosol profile
    during a period of relatively high uniformity and low atmospheric clarity.
    The two measurements are in good agreement up to $5$~km, in the region
    where aerosol attenuation has the greatest impact on FD observations.
    Despite the large differences in the operation, analysis, and viewing
    regions of the lidar and CLF, the optical measurements from the two
    instruments typically agree within their respective
    uncertainties~\cite{BenZvi:2007it}.  

    \subsubsection{Aerosol Optical Depth Uniformity}
    
\begin{figure}[ht]
      \begin{center}
        \includegraphics*[width=0.49\textwidth,clip]{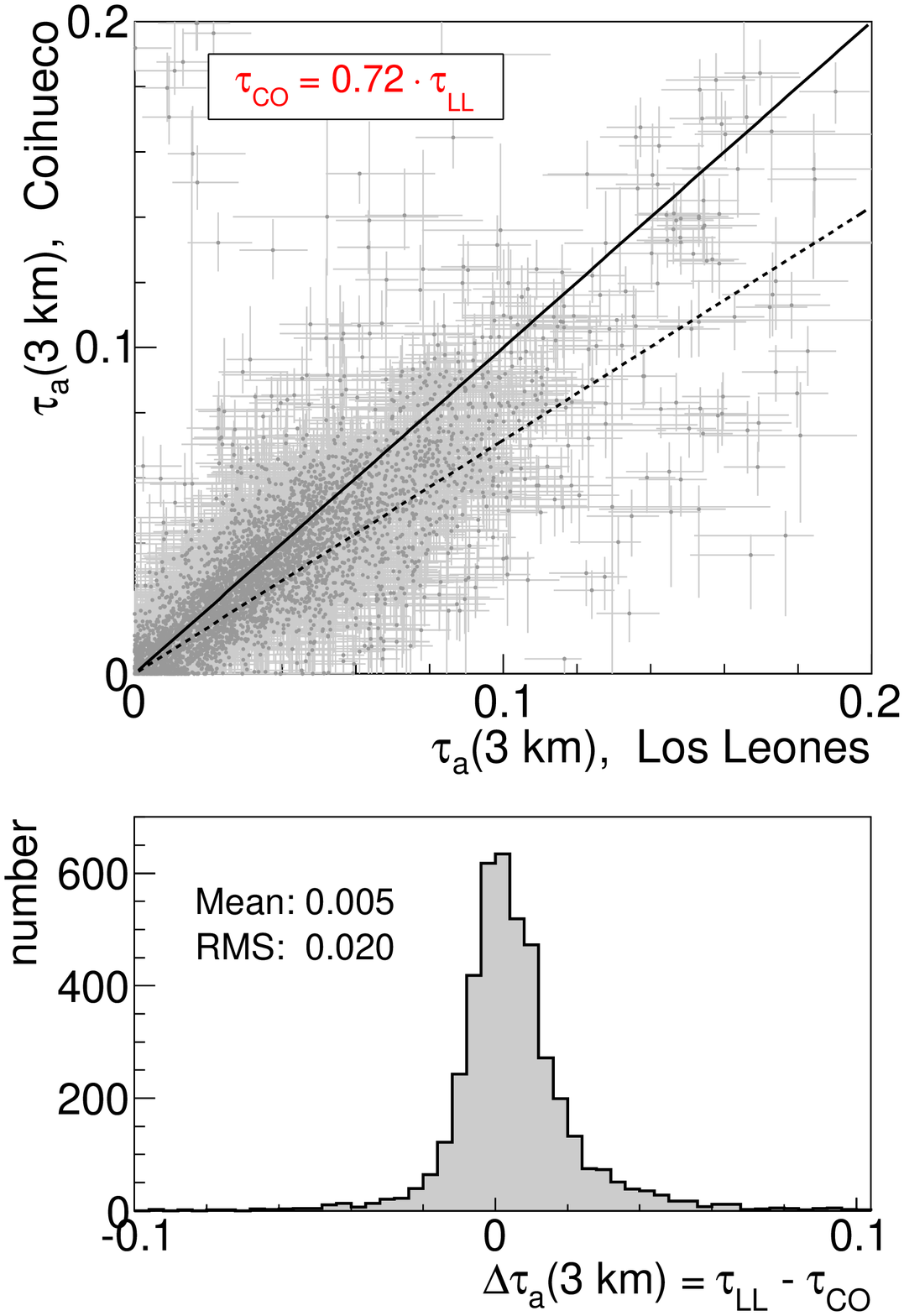}
        \includegraphics*[width=0.49\textwidth,clip]{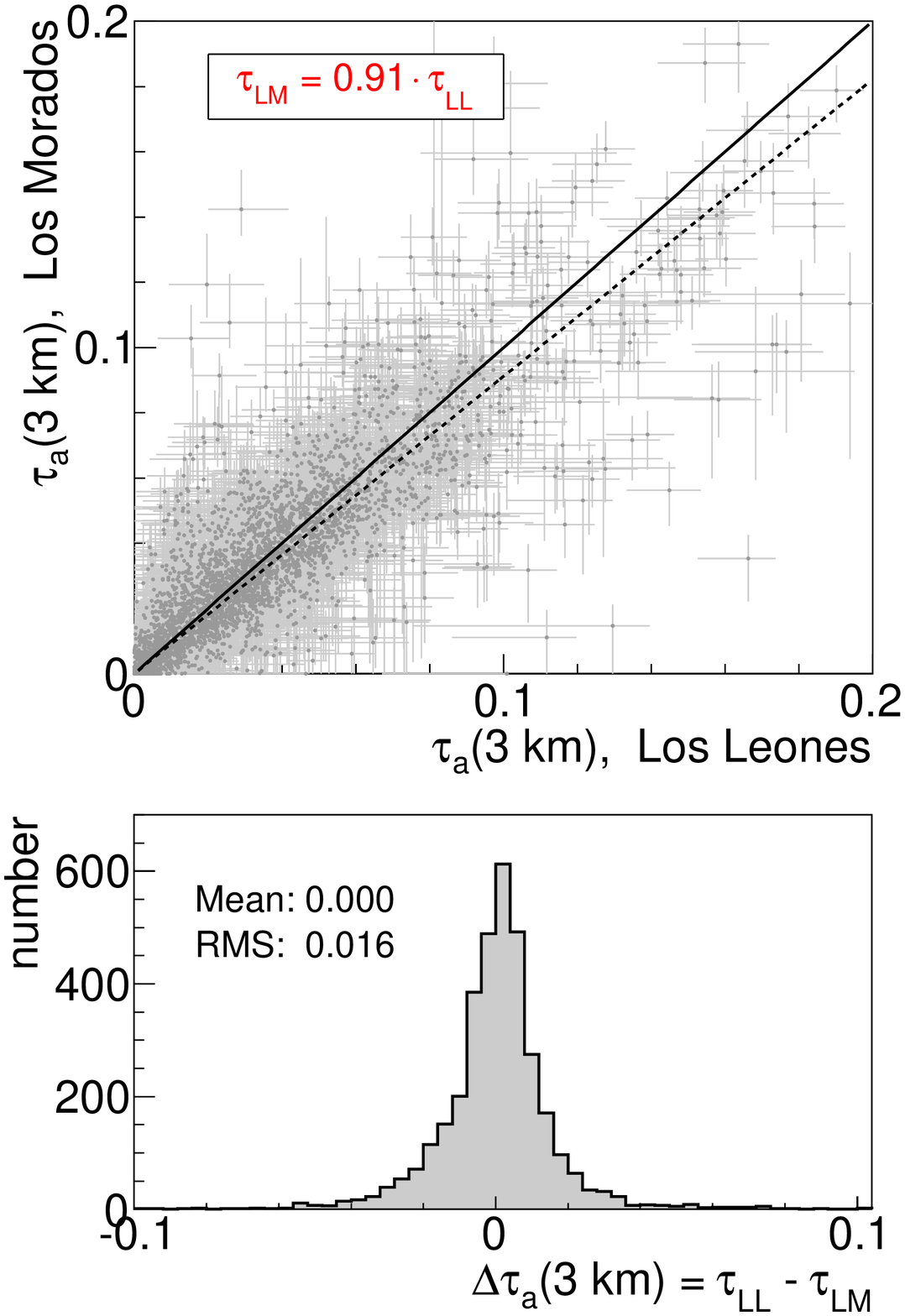}
        \caption{\label{fig:aerosol_uniformity} \slshape
          Comparison of the aerosol optical depths measured with CLF shots at
          Los Leones, Los Morados, and Coihueco.  The buildings at Los Leones
          and Los Morados are located on low hills at similar altitudes, while
          the Coihueco FD building is on a large hill $200$~m above the other
          sites.  The solid lines indicate equal optical depths at two
          sites, while the dotted lines show the best linear fits to the 
          optical depths.  The bottom panels show histograms of the differences
          between the optical depths.}
      \end{center}
    \end{figure}

    The FD building at Los Leones is located at an altitude of $1420$~m, on a
    hill about $15$~m above the surrounding plain, while the Coihueco site is
    on a ridge at altitude $1690$~m, a few hundred meters above the valley
    floor.  Since the distribution of aerosols follows the prevailing ground
    level rather than local irregularities, it is reasonable to expect that the
    aerosol optical depth between Coihueco and a fixed altitude will be
    systematically lower than the aerosol optical depth between Los Leones and
    the same altitude.  The data in fig.~\ref{fig:aerosol_uniformity} (left
    panel) support this expectation, and show that aerosol conditions differ
    significantly and systematically between these FD sites.  In contrast,
    optical depths measured at nearly equal altitudes, such as Los Leones and
    Los Morados (1420~m), are quite similar.

    Unlike for the molecular atmosphere, it is not possible to assume a
    horizontally uniform distribution of aerosols across the Observatory.  To
    handle the non-uniformity of aerosols between sites, the FD reconstruction
    divides the array into aerosol ``zones'' centered on the midpoints between
    the FD buildings and the CLF.  Within each zone, the vertical distribution
    of aerosols is treated as horizontally uniform by the reconstruction (i.e.,
    eq.~\eqref{eq:aerotrans} is applied).
    
  \subsection{Scattering Measurements}\label{subsec:aero_scat_measurements}
    
    Aerosol scattering is described by the phase function $P_a(\theta)$, and
    the hybrid reconstruction uses the functional form given in
    equation~\eqref{eq:aero_phase_function}.  As explained in
    Section~\ref{subsubsec:mol_aero_scattering}, the aerosol phase function for
    each hour must be determined with direct measurements of scattering in the
    atmosphere, which can be used to infer the backscattering and asymmetry
    parameters $f$ and $g$ of $P_a(\theta)$.

    \begin{figure}[ht]
      \begin{center}
        \includegraphics*[width=0.49\textwidth,clip]{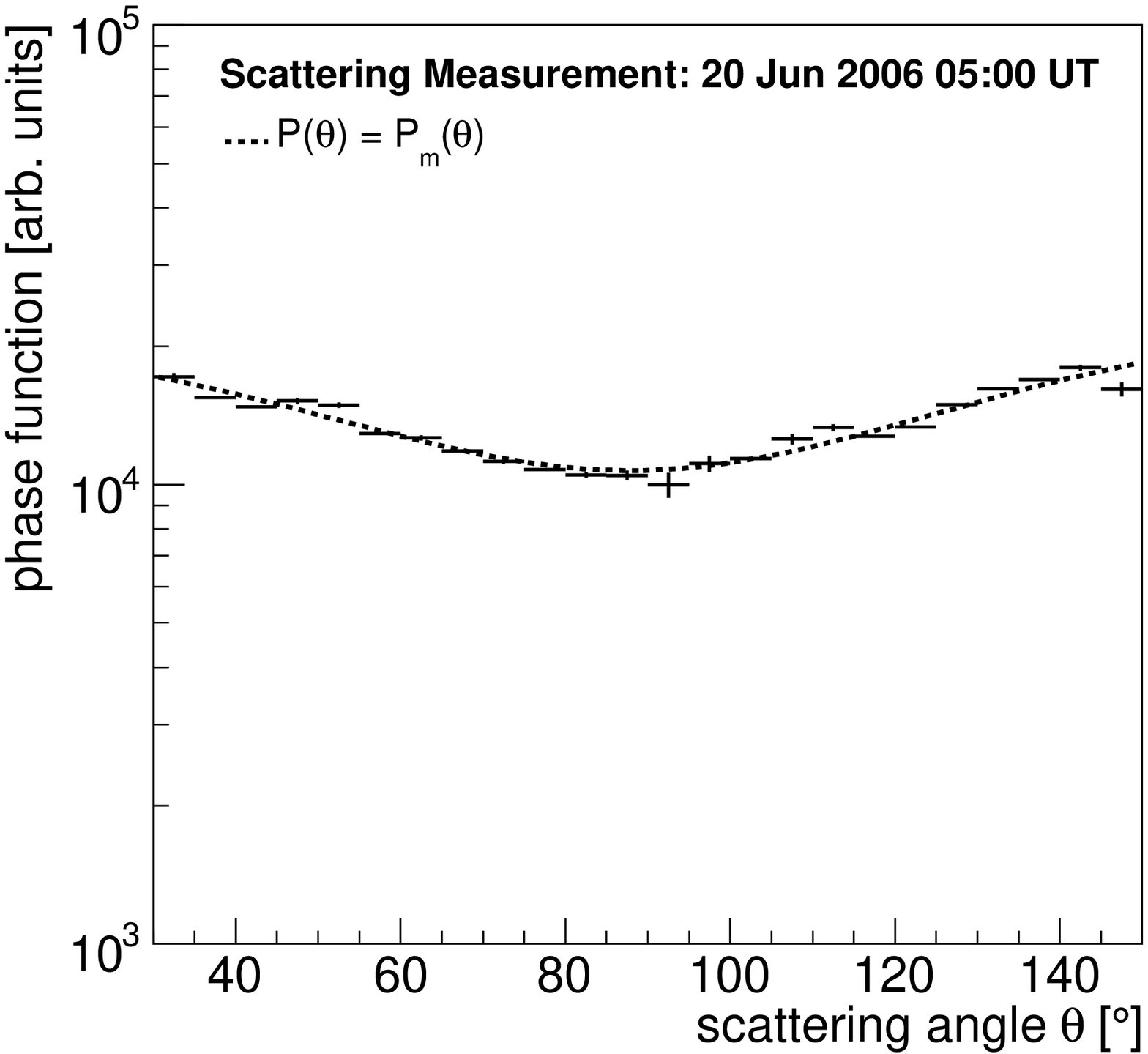}
        \includegraphics*[width=0.49\textwidth,clip]{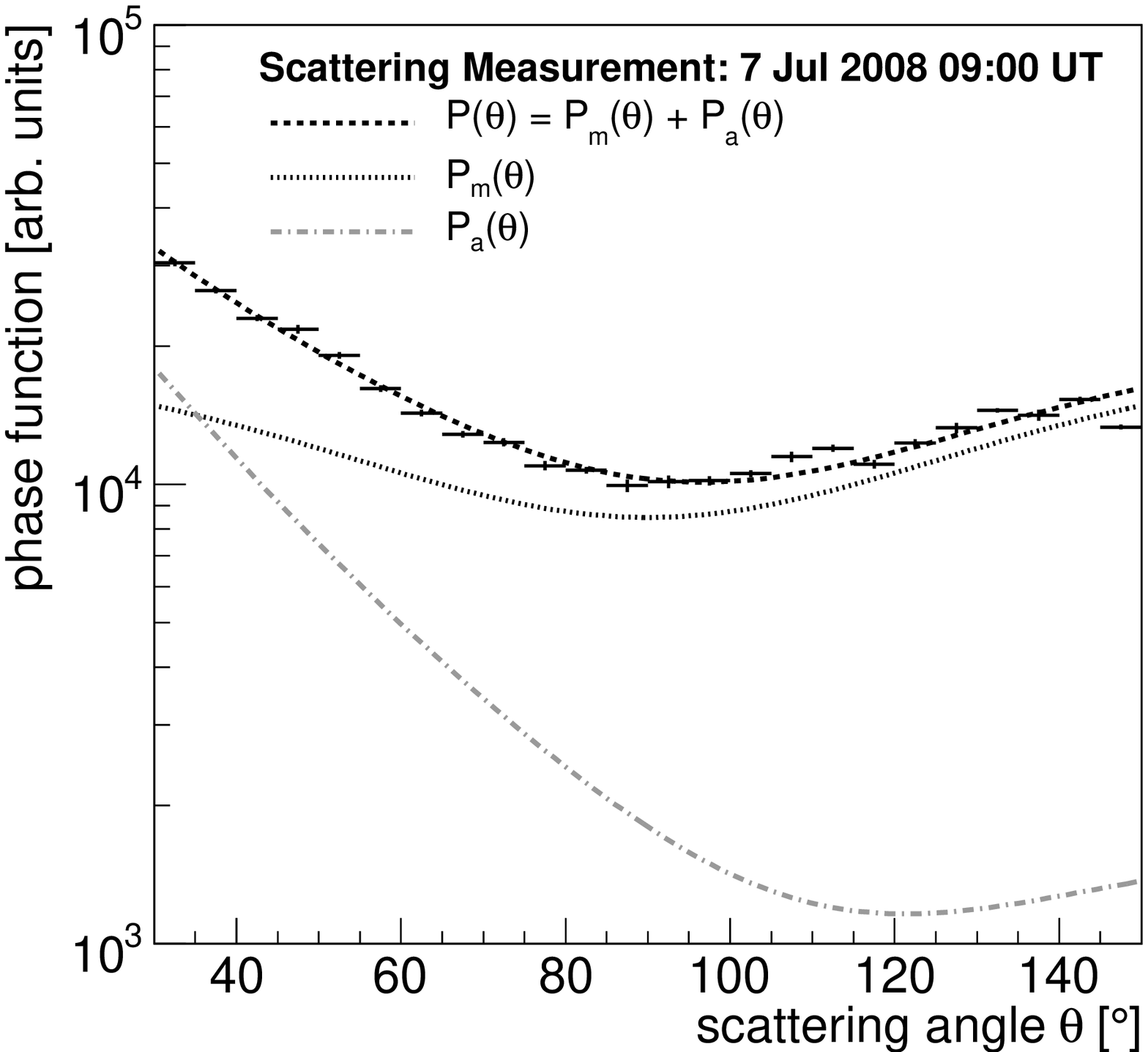}
        \caption{\label{fig:apf_fits} \slshape
          Light scattering measurements with the APF Xenon flasher at Coihueco.
          During a clear night (left), the observed phase function is symmetric
          due to the predominance of molecular scattering.  An asymmetric phase 
          function on a different night (right) indicates the presence of
          aerosols.}
      \end{center}
    \end{figure}
    
    At the Auger Observatory, these quantities are measured by two Aerosol
    Phase Function monitors, or APFs, located about $1$~km from the FD buildings
    at Coihueco and Los Morados~\cite{BenZvi:2007px}.  Each APF uses a
    collimated Xenon flash lamp to fire an hourly sequence of $350$~nm and
    $390$~nm shots horizontally across the FD field of view.  The shots are
    recorded during FD data acquisition, and provide a measurement of
    scattering at angles between $30^\circ$ and $150^\circ$.  A fit to the
    horizontal track seen by the FD is sufficient to determine $f$ and $g$.
    The APF light signal from two different nights is depicted in
    fig.~\ref{fig:apf_fits}, showing the total phase function fit and
    $P_a(\theta)$ after the molecular component has been subtracted.

    \begin{figure}[ht]
      \begin{center}
        \includegraphics*[width=\textwidth,clip]{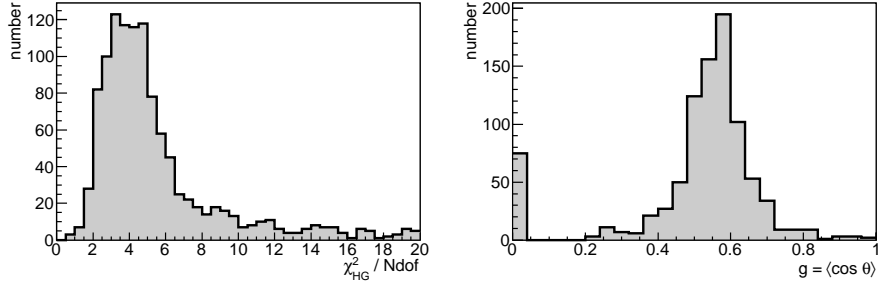}
        \caption{\label{fig:apfFitChi2G} \slshape
          Left: distribution of the figure of merit for fits of 
          equation~\eqref{eq:aero_phase_function} to APF measurements at
          Coihueco, June 2006 -- July 2008.  Right: distribution of the 
          asymmetry parameter $g$ measured at $350$~nm and $390$~nm at
          Coihueco.}
      \end{center}
    \end{figure}

    The phase function asymmetry parameter $g$ measured at Coihueco between
    June 2006 and July 2008 is shown in fig.~\ref{fig:apfFitChi2G}.  The value
    of $g$ was determined by fitting the modified Henyey-Greenstein function of
    eq.~\eqref{eq:aero_phase_function} to the APF data.  The reduced-$\chi^2$
    distribution for this fit, also shown in the figure, indicates that the
    Henyey-Greenstein function describes aerosol scattering in the FD
    reasonably well.  The measurements at Coihueco yield a site average
    $\langle g\rangle=0.56\pm0.10$ for the local asymmetry parameter, excluding
    clear nights without aerosol attenuation.  On clear (or nearly clear)
    nights, we estimate $g=0$ with an uncertainty of $0.2$.  The distribution
    of $g$ in \mal, a desert location with significant levels of sand and
    volcanic dust, is comparable to measurements reported in the literature for
    similar climates~\cite{2006JGRD..11105S04A}. 

    Approximately 900 hours of phase function data have been recorded with
    both APF monitors since June 2006.  The sparse data mean that it is not
    possible to use a true measurement of $P_a(\theta)$ for most FD events.
    Therefore, the Coihueco site average is currently used as the estimate of
    the phase function for all aerosol zones, for all cosmic ray events.  The
    systematic uncertainty introduced by this assumption will be explored in
    Section~\ref{subsubsec:aero_error_prop}.

  \subsection{Wavelength Dependence}\label{subsec:aero_wl_measurements}

    Measurements of the wavelength dependence of aerosol transmission are used
    to determine the \ang exponent $\gamma$ defined in
    equation~\eqref{eq:wavelength_dependence}.  At the \pao, observations of
    $\gamma$ are performed by two instruments: the Horizontal Attenuation
    Monitor, or HAM; and the (F/ph)otometric Robotic Telescope for Astronomical
    Monitoring, also known as FRAM~\cite{BenZvi:2007it,BenZvi:2007uj}.
    
    The HAM uses a high intensity discharge lamp located at Coihueco to provide
    an intense broad band light source for a CCD camera placed at Los Leones,
    about $45$~km distant.  This configuration allows the HAM to measure the
    total horizontal atmospheric attenuation across the Observatory.  To
    determine the wavelength dependence of the attenuation, the camera uses a
    filter wheel to record the source image at five wavelengths between $350$
    and $550$~nm.  By fitting the observed intensity as a function of
    wavelength, subtracting the estimated molecular attenuation, and assuming
    an aerosol dependence of the form of
    equation~\eqref{eq:wavelength_dependence}, it is possible to determine the
    \ang exponent $\gamma$ of aerosol attenuation.  During 2006 and 2007, the
    average exponent observed by the HAM was $\gamma=0.7$ with an RMS of $0.5$
    due to the non-Gaussian distribution of measurements~\cite{BenZvi:2007it}.
    The relatively small value of $\gamma$ suggests that \mal has a large
    component of coarse-mode aerosols.  This is consistent with physical
    expectations and other measurements in desert-like
    environments~\cite{Eck:1999,Kaskaoutis:2006}.
    
    Like the APF monitor data, the HAM and FRAM results are too sparse to use
    in the full reconstruction; therefore, during the FD reconstruction, the
    HAM site average for $\gamma$ is applied to all FD events in every aerosol
    zone. The result of this approximation is described in the next section.

%% file: hybridres.tex
\section{Impact of the Atmosphere on Accuracy of Reconstruction of Air Shower Parameters}\label{sec:aerosol_effects}

  The atmospheric measurements described in
  Sections~\ref{sec:molecular_effects} and \ref{sec:measurements} are fully
  integrated into the software used to reconstruct hybrid
  events~\cite{Allen:2008zz}.  The data are stored in multi-gigabyte MySQL
  databases and indexed by observation time, so that the atmospheric conditions
  corresponding to a given event are automatically retrieved during off-line
  reconstruction.  The software is driven by XML datacards that provide
  ``switches'' to study different effects on the
  reconstruction~\cite{Allen:2008zz}: for example, aerosol attenuation,
  multiple scattering, water vapor quenching, and other effects can be switched
  on or off while reconstructing shower profiles.  Propagation of
  atmospheric uncertainties is also available.

  In this section, we estimate the influence of atmospheric effects and the
  uncertainties in our knowledge of these effects on the reconstruction of
  hybrid events recorded between December 2004 and December 2008.  The data
  have been subjected to strong quality cuts to remove events contaminated with
  clouds\footnote{The presence of clouds distorts the observation of the shower
  profile as the UV light is strongly attenuated.  Clouds are also responsible
  for multiple scattering of the light.  Strong cuts on the shower profile
  shape can remove observations affected by clouds.}, as well as geometry cuts
  to eliminate events poorly viewed by the FD telescopes.  These cuts include:
  \begin{itemize}
    \item Gaisser-Hillas fit of the shower profile with $\chi^2/\text{NDF}<2.5$
    \item Gaps in the recorded light profile $<20\%$ of the length of the
          profile
    \item Shower maximum \xmax observed within the field of view of the FD
          telescopes
    \item Uncertainty in \xmax (before atmospheric corrections) $<40$~\gcmsq
    \item Relative uncertainty in energy (before atmospheric corrections) 
          $<20\%$
  \end{itemize}
  The cuts are the same as those used in studies of the energy
  spectrum~\cite{Schussler:2009,DiGiulio:2009} and \xmax
  distribution~\cite{Bellido:2009}.
  We first describe the effects of the molecular information on
  the determinations of energy and \xmax.  This is followed by a discussion of
  the impact of aerosol information on the measurement of these quantities.
  
  \subsection{Systematic Uncertainties due to the Molecular Atmosphere}\label{subsec:mol_uncertainties}

    \subsubsection{Monthly Models}\label{subsubsec:mmm_vary}

    The molecular transmission is determined largely by atmospheric pressure
    and temperature, as described in eq.~\eqref{eq:scatter_profile}.  For the
    purpose of reconstruction, these quantities are described by monthly
    molecular models generated using local radiosonde data.  Pressure and
    temperature also affect the fluorescence yield via collisional quenching,
    and this effect is included in the hybrid reconstruction. 

    \begin{figure}[ht]
      \begin{center}
        \includegraphics*[width=\textwidth,clip]{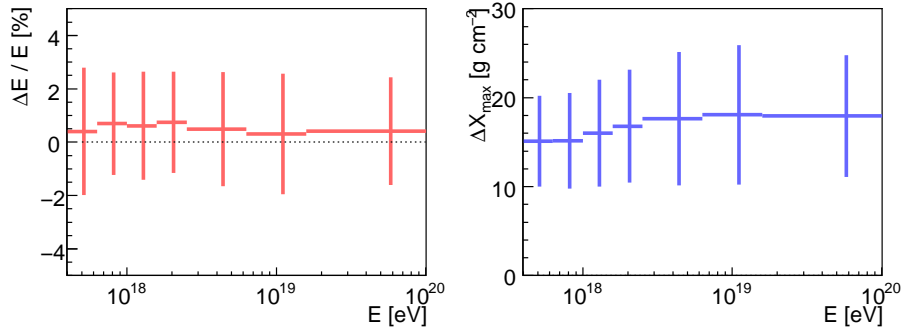}
        \caption{\label{fig:mmmVsUSStd} Comparison of hybrid events
        reconstructed using monthly balloon flights vs. the U.S. Standard
        Atmosphere (1976) profile model~\cite{USStdAtm:1976}.  Uncertainties
        denote the RMS spread.}
      \end{center}
    \end{figure}

    The importance of local atmospheric profile measurements is illustrated in
    fig.~\ref{fig:mmmVsUSStd}.  The hybrid data have been reconstructed using
    the monthly profile models described in Section~\ref{subsec:monthly_models}
    and compared to events reconstructed with the U.S. Standard
    Atmosphere~\cite{USStdAtm:1976}.  The values of \xmax determined using the
    molecular atmosphere described by the local monthly models are, on average,
    $15$~\gcmsq larger than the values obtained if the U.S. Standard Atmosphere
    is used.  This shift is energy-dependent because the average distance
    between shower tracks and the FD telescopes increases with energy.  It is
    clear that the U.S.  Standard Atmosphere is not an appropriate climate
    model for \mal; but even a local annual model would introduce seasonal
    shifts into the measurement of \xmax given the monthly variations observed
    in the local vertical depth profile (fig.~\ref{fig:monthly_profiles}).

    \begin{figure}[ht]
      \begin{center}
        \includegraphics*[width=\textwidth,clip]{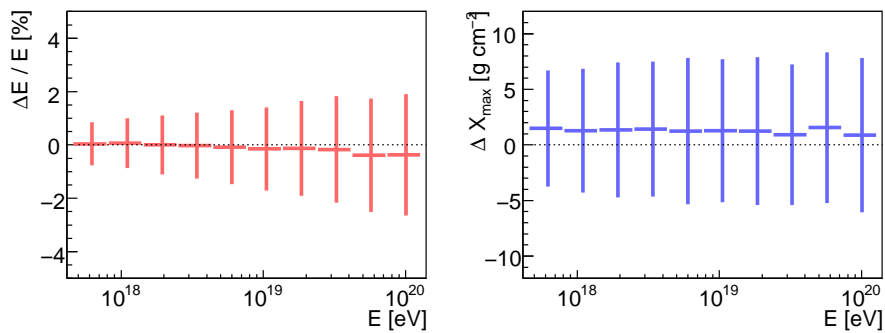}
        \caption{\label{fig:a07_mmmVsBalloon} Comparison of simulated events
        reconstructed with monthly average atmospheric profiles vs. profile
        measurements from 109 cloud-free balloon flights.  The dotted lines
        indicate the reference for the 109 balloon flights; the uncertainties
        indicate the RMS spread.}
      \end{center}
    \end{figure}

    When the monthly models are used, some systematic uncertainties are
    introduced into the reconstruction due to atmospheric variations that occur
    on timescales shorter than one month.  To investigate this effect, we
    compare events reconstructed with monthly models vs. local radio soundings.
    A set of 109 cloud-free, night-time balloon profiles was identified using
    the cloud camera database.  The small number of soundings requires the use
    of simulated events, so we simulated an equal number of proton and iron
    showers between $10^{17.5}$~eV and $10^{20}$~eV, reconstructed them with
    monthly and radiosonde profiles, and applied standard cuts to the simulated
    dataset.  The radiosonde profiles were weighted in the simulation to
    account for seasonal biases in the balloon launch rate.

    The difference between monthly models and balloon measurements is indicated
    in fig.~\ref{fig:a07_mmmVsBalloon}.  The use of the models introduces
    rather small shifts into the reconstructed energy and \xmax, though there
    is an energy-dependent increase in the RMS of the measured energies from
    $0.8\%$ to $2.0\%$ over the simulated energy range.  The systematic shift
    in \xmax is about $2$~\gcmsq over the full energy scale, with an RMS of
    about $8$~\gcmsq.  We interpret the RMS spread as the decrease in
    the resolution of the hybrid detector due to variations in the atmospheric
    conditions within each month.

    \subsubsection{Combined Effects of Quenching and Atmospheric
    Variability}\label{subsubsec:quench_vary}

    The simulations described in Section~\ref{subsubsec:mmm_vary} used an air
    fluorescence model that does not correct the fluorescence emission for
    weather-dependent quenching.  Recent estimates of quenching due to
    water vapor~\cite{Morozov:2005,Waldenmaier:2007am} and the $T$-dependence
    of the \nitrogen-\nitrogen and \nitrogen-\oxygen collisional cross
    sections~\cite{Keilhauer:2008sy} allow for detailed studies of their effect
    on the production of fluorescence light.

    \begin{figure}[ht]
      \begin{center}
        \includegraphics*[width=\textwidth,clip]{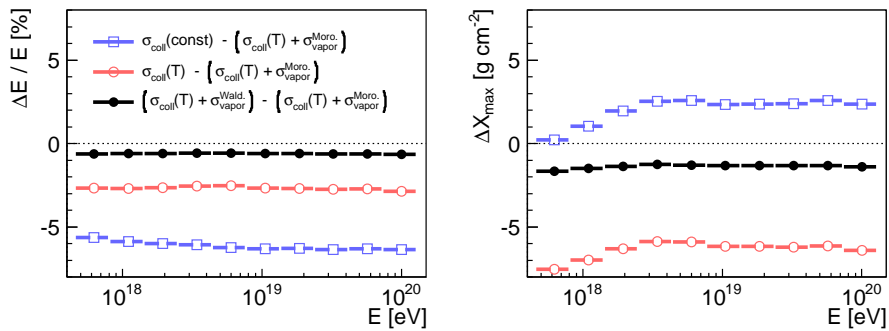}
        \caption{\label{fig:k08Comparison} The effect of water vapor quenching
        and the $T$-dependent \nitrogen-\nitrogen and \nitrogen-\oxygen
        collisional cross sections $\sigma_\text{coll}(T)$ on the reconstructed
        energy and \xmax of simulated showers.  The reference (dotted line)
        corresponds to showers reconstructed with the fluorescence model of
        Keilhauer et al.~\cite{Keilhauer:2008sy}, which includes $T$-dependent
        cross sections, and the vapor quenching model of Morozov et al.
        $(\sigma^\text{Moro.}_\text{vapor})$~\cite{Morozov:2005}.  The markers
        correspond to different combinations of quenching effects and vapor
        quenching models.  See the text for a detailed explanation.}
      \end{center}
    \end{figure}
    
    We have applied the two quenching effects to simulated showers in
    various combinations using $p$, $T$, and $u$ from the monthly model
    profiles (see fig.~\ref{fig:k08Comparison}).  As different quenching
    effects and models were ``switched on'' in the reconstruction, the showers
    were compared to a reference reconstruction that used $T$-dependent
    collisional cross sections and the vapor quenching model of Morozov et
    al.~\cite{Morozov:2005}.  We have considered the following three cases:
    \begin{enumerate}
      \item In the first case, all quenching corrections were omitted (open blue
            squares in fig.~\ref{fig:k08Comparison}).  The result is a
            $5.5\%$ underestimate in shower energy and a $2$~\gcmsq overestimate
            in \xmax with respect to the reference model.
      \item In the second case, temperature corrections to the collisional
            cross section were included, but water vapor quenching was not (open
            red circles in the figure).  Without vapor quenching, the energy
            is systematically underestimated by $3\%$, and \xmax is
            underestimated by $6-7$~\gcmsq with respect to the reference model.
      \item In the third case, all corrections were included, but the vapor
            quenching model of Waldenmaier et al.~\cite{Waldenmaier:2007am} was
            used (closed black circles).  The resulting systematic differences
            are \relE$=0.5\%$ and \dxmax=$2$~\gcmsq.
    \end{enumerate}
    We observe that once water vapor quenching is applied, the particular
    choice of quenching model has a minor influence.  In addition, there is a
    small total shift in \xmax due to the offsetting effects of the
    $T$-dependent cross sections, which are important at high altitudes, and
    the effect of water vapor, which is important at low altitudes.  The
    compensation of these two effects leaves the longitudinal profiles of
    showers relatively undistorted.
    
    \begin{figure}[ht]
      \begin{center}
        \includegraphics*[width=\textwidth,clip]{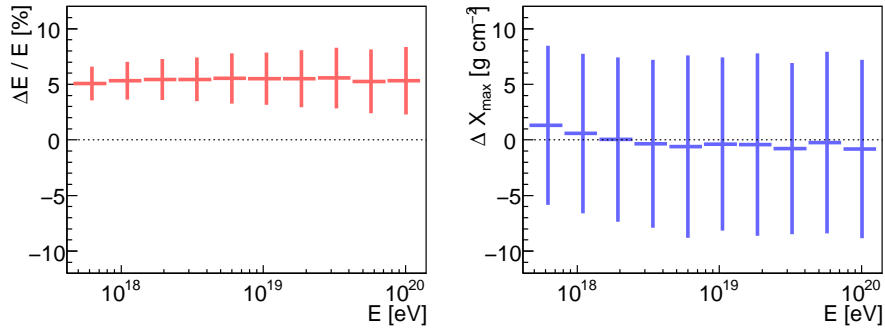}
        \caption{\label{fig:k08a07_mmmVsBalloon}
          Comparison of simulated events to determine the combined effects of
          atmospheric variability and quenching corrections.  The data were
          reconstructed in two sets: using monthly profiles plus a fluorescence
          model without quenching corrections~\cite{Ave:2007xh}; and using 
          local radiosonde profiles plus a fluorescence model with water vapor
          quenching and $T$-dependent collisional cross
          sections~\cite{Keilhauer:2008sy}.}
      \end{center}
    \end{figure}

    In fig.~\ref{fig:k08a07_mmmVsBalloon}, we plot the combined effects of
    atmospheric variability around the monthly averages and the quenching
    corrections.  Simulated showers were reconstructed with two settings:
    monthly average profile models and no quenching corrections; and cloud-free
    radiosonde profiles with water vapor quenching and $T$-dependent
    collisional cross sections.  The reconstructed energy is increased by
    $5\%$, on average, and, comparing figs.~\ref{fig:a07_mmmVsBalloon} and
    \ref{fig:k08a07_mmmVsBalloon}, we see that the quenching corrections are
    dominating systematic uncertainties due to the use of monthly models.  For
    \xmax, the systematic effects of the monthly models offset the quenching
    corrections.  The spread of the combined measurements increases with
    energy, such that the RMS in energy increases from $1.5\%-3.0\%$, and the
    RMS in \xmax increases from $7.2-8.4$~\gcmsq.

  \subsection{Uncertainties due to Aerosols}

  For a complete understanding of the effects of aerosols on the
  reconstruction, several investigations are of interest:
  \begin{enumerate}
    \item A test of the effect of aerosols on the reconstruction, compared to
          the use of a pure molecular atmosphere.
    \item A test of the use of aerosol measurements, compared to a simple
          parameterization of average aerosol conditions.
    \item The propagation of measurement uncertainties in \vaod, $\gamma$, $f$,
          and $g$ in the FD reconstruction, and in particular their effect on
          uncertainties in energy and \xmax.
    \item A test of the horizontal uniformity of aerosol layers within a
          zone.
  \end{enumerate}

  \subsubsection{The Comparison of Aerosol Measurements with a Pure Molecular Atmosphere}

\begin{figure}[ht]
    \begin{center}
      \includegraphics*[width=\textwidth,clip]{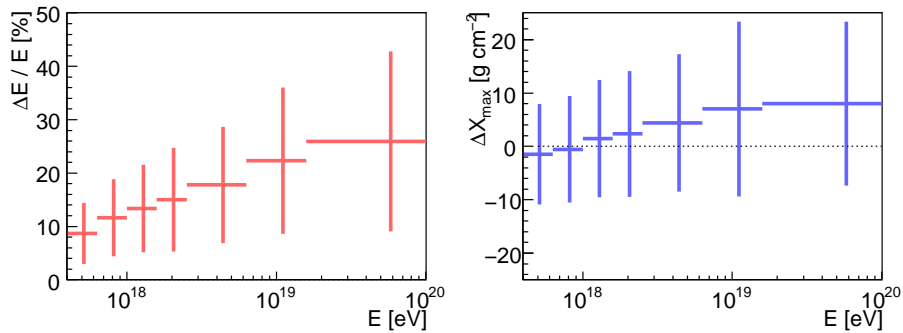}
      \caption{\label{fig:aero_vs_mol} \slshape 
        Comparison of hybrid events reconstructed with hourly CLF aerosol
        measurements vs. no aerosol correction (i.e., purely molecular
        transmission).  Uncertainties indicate the RMS spread for each energy.}
    \end{center}
  \end{figure}

  We have compared the reconstruction of hybrid showers using hourly on-site
  aerosol measurements with the same showers reconstructed using a purely
  molecular atmosphere (fig.~\ref{fig:aero_vs_mol}).  Neglecting the presence
  of aerosols causes an $8\%$ underestimate in energy at the lower energies.
  This underestimate increases to $25\%$ at the higher energies.  Moreover, the
  distribution of shifted energies contains a long tail: $20\%$ of all showers
  have an energy correction $>20\%$; $7\%$ of showers are corrected by $>30\%$;
  and $3\%$ of showers are corrected by $>40\%$.  The systematic shift in \xmax
  ranges from $-1$~\gcmsq at low energies to almost $10$~\gcmsq at high
  energies, with an RMS of $10-15$~\gcmsq.
  
  \subsubsection{The Comparison of Aerosol Measurements with an Average Parameterization}

  Aerosols clearly play an important role in the transmission and scattering of
  fluorescence light, but it is natural to ask if hourly measurements of
  aerosol conditions are necessary, or if a fixed average aerosol
  parameterization is sufficient for air shower reconstruction.  

\begin{figure}[ht]
    \begin{center}
      \includegraphics[width=\textwidth]{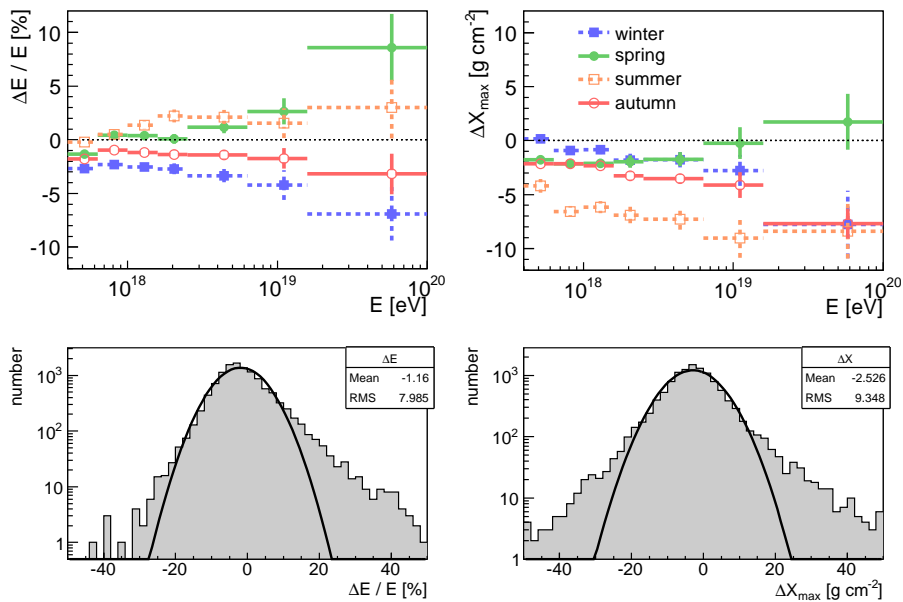}
      \caption{\label{fig:aerosol_seasonal} \slshape
        Top: systematic shifts in the hybrid reconstruction of shower energy and
        \xmax caused by the use of average aerosol conditions rather than
        hourly measurements (indicated by dotted lines).  The mean
        $\Delta E/E$ and $\Delta X_\text{max}$ per energy bin, plotted with
        uncertainties on the means, are arranged by their occurrence in
        austral winter, spring, summer, and autumn. Bottom: distributions of the
        differences in energy and \xmax, shown with Gaussian fits.}
    \end{center}
  \end{figure}

  We can test the sufficiency of average aerosol models by comparing the
  reconstruction of hybrid events using hourly weather data against the
  reconstruction using an average profile of the aerosol optical depth in \mal.
  The average profile was constructed using CLF data, and the differences in
  the reconstruction between this average model and the hourly data are shown
  in fig.~\ref{fig:aerosol_seasonal}, where $\Delta E/E$ and $\Delta
  X_\text{max}$ are grouped by season.
  
  Due to the relatively good viewing conditions in \mal during austral winter
  and fall, and poorer atmospheric clarity during the spring and summer, the
  shifts caused by the use of an average aerosol profile exhibit a strong
  seasonal dependence.  The shifts also exhibit large tails and are
  energy-dependent.  For example, $\Delta E/E$ nearly doubles during the fall,
  winter, and spring, reaching
  $-7\%$ (with an RMS of $15\%$) during the winter.  The range of seasonal mean
  offsets in \xmax is $+2$~\gcmsq to $-8$~\gcmsq (with an RMS of $15$~\gcmsq),
  and the offsets depend strongly on the shower energy.
  
  \subsubsection{Propagation of Uncertainties in Aerosol Measurements}\label{subsubsec:aero_error_prop}

  Uncertainties in aerosol properties will cause over- or under-corrections of
  recorded shower light profiles, particularly at low altitudes and low
  elevation angles.  On average, systematic overestimates of the aerosol
  optical depth will lead to an over-correction of scattering losses and an
  overestimate of the shower light flux from low altitudes; this will increase
  the shower energy estimate and push the reconstructed \xmax deeper into the
  atmosphere.  Systematic underestimates of the aerosol optical depth should
  have the opposite effect.
  
  \begin{figure}[ht]
    \begin{center}
      \includegraphics*[width=\textwidth,clip]{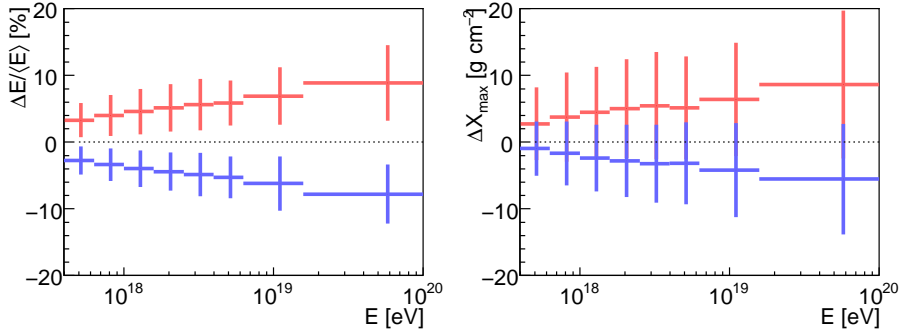}
      \caption{\label{fig:aero_uncertainty} \slshape Shifts in the
        reconstruction of energy and \xmax when the aerosol optical depth is 
        varied by its $+1\sigma$ systematic uncertainty (red points) and 
        $-1\sigma$ systematic uncertainty (blue points).  The dotted line
        corresponds to the central aerosol optical depth measurement.  The
        uncertainty bars correspond to the sample RMS in each energy bin.}
    \end{center}
  \end{figure}

  The primary source of uncertainty in aerosol transmission comes from the
  aerosol optical depth~\cite{Prouza:2007ur} estimated using vertical
  CLF laser shots.  The uncertainties in the hourly CLF optical depth profiles
  are dominated by systematic detector and calibration effects, and smoothing
  of the profiles makes the optical depths at different altitudes highly
  correlated.  Therefore, a reasonable estimate of the systematic uncertainty
  in energy and \xmax can be obtained by shifting the full optical depth
  profiles by their uncertainties and estimating the mean change in the
  reconstructed energy and \xmax.

  This procedure was done using hybrid events recorded by telescopes at Los
  Leones, Los Morados, and Coihueco, and results are shown in
  fig.~\ref{fig:aero_uncertainty}.  The energy dependence of the uncertainties
  mainly arises from the distribution of showers with distance: low-energy
  showers tend to be observed during clear viewing conditions and within
  $10$~km of the FD buildings, reducing the effect of the transmission
  uncertainties on the reconstruction; and high-energy showers can be observed
  in most aerosol conditions (up to a reasonable limit) and are observed at
  larger distances from the FD.  The slight asymmetry in the shifts is due to
  the asymmetric uncertainties of the optical depth profiles.

  By contrast to the corrections for the optical depth of the aerosols, the
  uncertainties that arise from the wavelength dependence of the aerosol
  scattering and of the phase function are relatively unimportant for the
  systematic uncertainties in shower energy and \xmax.  By reconstructing
  showers with average values of the \ang coefficient and the phase function
  measured at the Observatory, and comparing the results to showers
  reconstructed with the $\pm1~\sigma$ uncertainties in these measurements, we
  find that the wavelength dependence and phase function contribute $0.5\%$ and
  $1\%$, respectively, to the uncertainty in the energy, and $\sim2$~\gcmsq to
  the systematic uncertainty in \xmax \cite{Prouza:2007ur}.  Moreover, the
  uncertainties are largely independent of shower energy and distance.

  \subsubsection{Evaluation of the Horizontal Uniformity of the Atmosphere}

  The non-uniformity of the molecular atmosphere, discussed in
  Section~\ref{subsec:horizontal_molecular}, is very minor and introduces
  uncertainties $<1\%$ in shower energies and about $1$~\gcmsq in \xmax.
  Non-uniformities in the horizontal distribution of aerosols may also be
  present, and we expect these to have an effect on the reconstruction.  For
  each FD building, the vertical CLF laser tracks only probe the atmosphere
  along one light path, but the reconstruction must use this single aerosol
  profile across the azimuth range observed at each site.  In general, the
  assumption of uniformity within an aerosol zone is reasonable, though the
  presence of local inhomogeneities such as clouds, fog banks, and sources of
  dust and smoke may render it invalid.

  The assumption of uniformity can be partially tested by comparing data
  reconstructed with different aerosol zones around each eye: for example,
  reconstructing showers observed at Los Leones using aerosol data from the Los
  Leones and Los Morados zones.

  \begin{figure}[ht]
    \begin{center}
      \includegraphics[width=\textwidth]{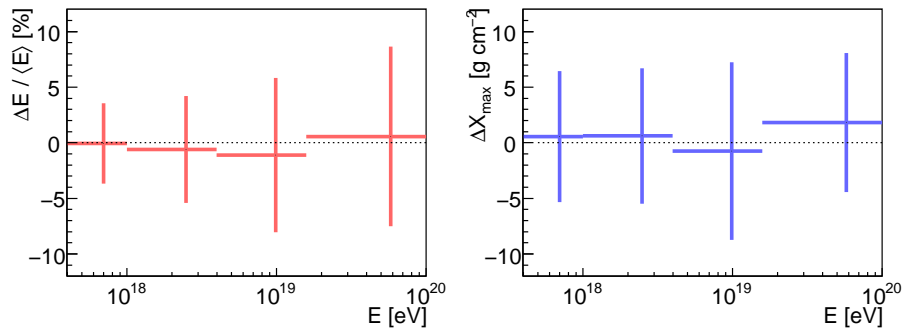}
      \caption{\label{fig:aero_uniformity} \slshape Shifts in the estimated
               shower energy and \xmax when data from the FD buildings at Los
               Morados and Los Leones (dotted line) are reconstructed with
               swapped aerosol zones.  The values give an approximate estimate
               of the systematic uncertainty due to aerosol non-uniformities
               across the detector.  The uncertainties correspond to the 
               sample RMS in each energy bin.}
    \end{center}
  \end{figure}

  Data from Los Leones and Los Morados were reconstructed using aerosol
  profiles from both zones, and the resulting profiles are compared in
  fig.~\ref{fig:aero_uniformity}.  The mean shifts \relE and \dxmax are
  relatively constant with energy: \relE$=0.5\%$, and \dxmax is close to zero.
  The distributions of \relE and \dxmax are affected by long tails, with the
  RMS in \relE growing with energy from $3\%$ to $8\%$.  For \dxmax, the RMS
  for all energies is about 6~\gcmsq.  

  \subsection{Corrections for Multiple Scattering}\label{subsec:ms_correction}
  
  Multiply-scattered light, if not accounted for in the reconstruction, will
  lead to a systematic overestimate of shower energy and \xmax.  This is
  because multiple scattering shifts light into the FD field of view that would
  otherwise remain outside the shower image.  A na\"{i}ve reconstruction will
  incorrectly identify multiply-scattered photons as components of the direct
  fluorescence/Cherenkov and singly-scattered Cherenkov signals, leading to an
  overestimate of the Cherenkov-fluorescence light production used in the
  calculation of the shower profile.  The mis-reconstruction of \xmax is
  similar to what occurs in the case of overestimated optical depths: not
  enough scattered light is removed from the low-altitude tail of the shower
  profile, causing an overestimate of $dE/dX$ in the deep part of the profile.

\begin{figure}[ht]
    \begin{center}
      \includegraphics[width=\textwidth]{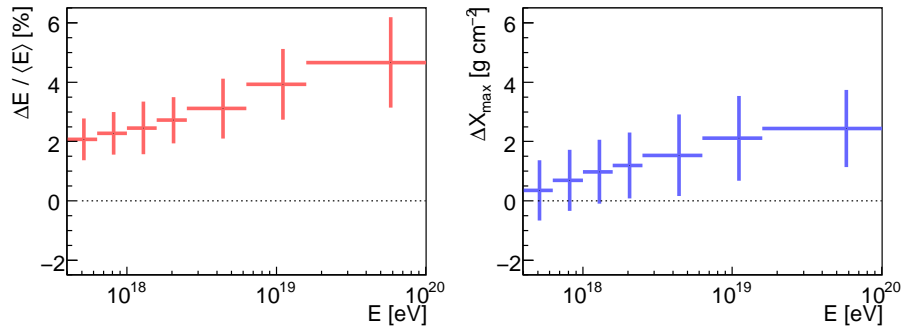}
      \caption{\label{fig:ms_corrections} \slshape Overestimates of shower
               energy (left) and \xmax (right) due to lack of multiple
               scattering corrections in the hybrid reconstruction.  The dotted
               lines correspond to a reconstruction with multiple scattering
               enabled.  The uncertainties correspond to the sample RMS in
               each energy bin.}
    \end{center}
  \end{figure}

  The parameterizations of multiple scattering due to
  Roberts~\cite{Roberts:2005xv} and Pekala et al.~\cite{Pekala:2009fe} have
  been implemented in the hybrid event reconstruction.  The predictions from
  both analyses are that the scattered light fraction in the shower image will
  increase with optical depth, so that distant high-energy showers will be most
  affected by multiple scattering.  A comparison of showers reconstructed with
  and without multiple scattering (fig.~\ref{fig:ms_corrections}) verifies that
  the shift in the estimated energy doubles from $2\%$ to nearly $5\%$ as the
  shower energy (and therefore, average shower distance to the FD) increases.
  The systematic error in the shower maximum is also consistent with the
  overestimate of the light signal that occurs without multiple scattering
  corrections.

\begin{figure}[ht]
    \begin{center}
      \includegraphics[width=\textwidth]{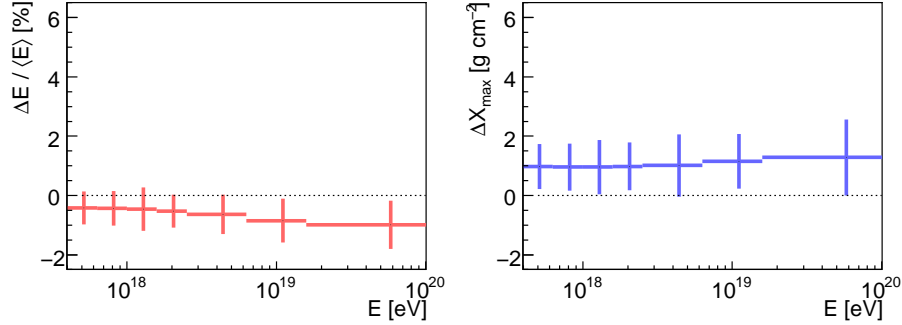}
      \caption{\label{fig:ms_implement} \slshape Systematic differences in
               shower energy (left) and \xmax (right) for events
               reconstructed using the multiple scattering corrections of
               Roberts~\cite{Roberts:2005xv} (dotted lines) and Pekala et 
               al.~\cite{Pekala:2009fe}.}
    \end{center}
  \end{figure}

  The multiple scattering corrections due to Roberts and Pekala et al. give
  rise to small differences in the reconstructed energy and \xmax.  As shown in
  fig.~\ref{fig:ms_implement}, the two parameterizations differ in the energy
  correction by $<1\%$, and there is a shift of $1$~\gcmsq in \xmax for all
  energies.  These values provide an estimate of the systematic uncertainties
  due to multiple scattering which remain in the reconstruction.
  
  \subsection{Summary}\label{subsec:hrec_unc_summary}

  \begin{table}[ht]
    \def\arraystretch{1.05}
    \begin{center}
      \footnotesize{
      \begin{tabular}{|l|c|c|c|c|c|}
      \hline
      \multicolumn{6}{|c|}{
        \rule[-3mm]{0mm}{8mm}\bfseries Systematic Uncertainties}\\
      \hline
      \multirow{2}{*}{Source} & \multirow{2}{*}{$\log{(E/\text{eV})}$} 
                      & \relE & RMS$(\Delta E/E)$
                      & \dxmax & RMS$(X_\text{max})$ \\
      & & (\%) & (\%) & (\gcmsq) & (\gcmsq) \\
      \hline
      \multicolumn{6}{|c|}{\slshape Molecular Light Transmission and Production} \\
      \hline
      Horiz. Uniformity & $17.7-20.0$ & 1 & 1 & 1 & 2 \\
      \hline
      Quenching Effects & $17.7-20.0$ & $+5.5$ & \multirow{2}{*}{$1.5-3.0$}
                            & $-2.0$ & \multirow{2}{*}{$7.2-8.4$} \\
      $p$, $T$, $u$ Variability & $17.7-20.0$ & $-0.5$ & & $+2.0$ & \\
      \hline
      \multicolumn{6}{|c|}{\slshape Aerosol Light Transmission} \\
      \hline
      \multirow{3}{*}{Optical Depth}
          & $<18.0$ & $+3.6$, $-3.0$ & $1.6\pm1.6$ & $+3.3$, $-1.3$ & $3.0\pm3.0$ \\
          & $18.0-19.0$ & $+5.1$, $-4.4$ & $1.8\pm1.8$ & $+4.9$, $-2.8$ & $3.7\pm3.7$ \\
          & $19.0-20.0$ & $+7.9$, $-7.0$ & $2.5\pm2.5$ & $+7.3$, $-4.8$ & $4.7\pm4.7$ \\
      \hline
      $\lambda$-Dependence & $17.7-20.0$ & 0.5 & 2.0 & 0.5 & 2.0 \\
      \hline
      Phase Function & $17.7-20.0$ & 1.0 & 2.0 & 2.0 & 2.5 \\
      \hline
      \multirow{3}{*}{Horiz. Uniformity}
        & $<18.0$ & 0.3 & 3.6 & 0.1 & 5.7 \\
        & $18.0-19.0$ & 0.4 & 5.4 & 0.1 & 7.0 \\
        & $19.0-20.0$ & 0.2 & 7.4 & 0.4 & 7.6 \\
      \hline
      \multicolumn{6}{|c|}{\slshape Scattering Corrections} \\
      \hline
      \multirow{3}{*}{Mult. Scattering}
        & $<18.0$ & 0.4 & 0.6 & 1.0 & 0.8 \\
        & $18.0-19.0$ & 0.5 & 0.7 & 1.0 & 0.9 \\
        & $19.0-20.0$ & 1.0 & 0.8 & 1.2 & 1.1 \\
      \hline
      \end{tabular}
      }
      \caption{\slshape Systematic uncertainties in the hybrid reconstruction
               due to atmospheric influences on light transmission or
               production.}
      \label{table:uncertainties}
    \end{center}
    \def\arraystretch{1.5}
  \end{table}

  Table~\ref{table:uncertainties} summarizes our estimate of the impact of the
  atmosphere on the energy and \xmax measurements of the hybrid detector of the
  \pao.  Aside from large quenching effects due to missing quenching
  corrections in the reconstruction, the systematic uncertainties are currently
  dominated by the aerosol optical depth: $4-8\%$ for shower energy, and about
  $4-8$~\gcmsq for \xmax.  This list of uncertainties is similar to that
  reported in~\cite{Prouza:2007ur}, but now includes an explicit statement of
  the multiple scattering correction\footnote{Note that in previous
  publications, this correction has been absorbed into a more general $10\%$
  systematic uncertainty due to reconstruction
  methods~\cite{Dawson:2007di,Abraham:2008ru}.}.

  The RMS values in the table can be interpreted as the spread in measurements
  of energy and \xmax due to current limitations in the atmospheric monitoring
  program.  For example, the uncertainties due to the variability of $p$, $T$,
  and $u$ are caused by the use of monthly molecular models in the
  reconstruction rather than daily measurements, while uncertainties due to the
  horizontal non-uniformity of aerosols are due to limited spatial sampling of
  the full atmosphere.  Note that the RMS values listed for the aerosol optical
  depth are due to a mixture of systematic and statistical uncertainties; we
  have estimated these contributions conservatively by expressing the RMS as a
  central value with large systematic uncertainties.  The combined values from
  all atmospheric measurements are, approximately, RMS(\relE)$\approx5\pm1\%$
  to $9\pm1\%$ as a function of energy, and RMS(\xmax)$\approx11\pm1$~\gcmsq to
  $13\pm1$~\gcmsq.  In principle, the RMS can be reduced by improving the
  spatial resolution and timing of the atmospheric monitoring data.  Such
  efforts are underway, and are described in Section~\ref{sec:future}.

%% file: futuredevel.tex
\section{Additional Developments}\label{sec:future}

  We have estimated the uncertainties in shower energy and \xmax due to
  atmospheric transmission, but we have not discussed the impact of clouds on
  the hybrid reconstruction, which violate the horizontal uniformity assumption
  described in section~\ref{subsec:weath_trans}.  A full treatment of this
  issue will be the subject of future technical publications, but here we
  summarize current efforts to understand their effect on the hybrid data.

  \subsection{Cloud Measurements}\label{subsec:clouds}

\begin{figure}[ht]
    \begin{center}
      \includegraphics*[width=0.55\textwidth,clip]{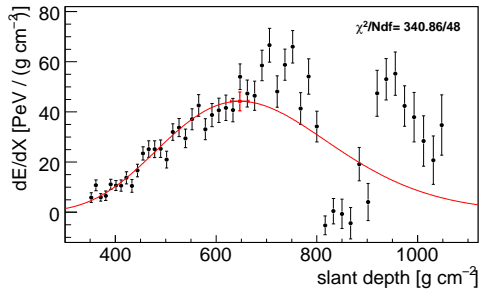}
      \caption{\label{fig:profile_with_cloud} \slshape Shower light profile
               with a large gap due to the presence of an intervening cloud.}
    \end{center}
  \end{figure}

  Cloud coverage has a major influence on the reconstruction of air showers,
  but this influence can be difficult to quantify.  Clouds can block the
  transmission of light from air showers, as shown in
  Figure~\ref{fig:profile_with_cloud}, or enhance the observed light flux due
  to multiple scattering of the intense Cherenkov light beam.  They may occur
  in optically thin layers near the top of the troposphere, or in thick banks
  which block light from large parts of the FD fiducial volume.  The
  determination of the composition of clouds is nontrivial, making \textsl{a
  priori} estimates of their scattering properties unreliable.

  Due to the difficulty of correcting for the transmission of light through
  clouds, it is prudent to remove cloudy data using hard cuts on the shower
  profiles.  But because clouds can reduce the event rate from different parts
  of a fluorescence detector, they also have an important effect on the
  aperture of the detector as used in the determination of the spectrum from
  hybrid data~\cite{Schussler:2009}.  Therefore, it is necessary to estimate
  the cloud coverage at each FD site as accurately as possible.

\begin{figure}[ht]
    \begin{center}
      \includegraphics*[width=\textwidth,clip]{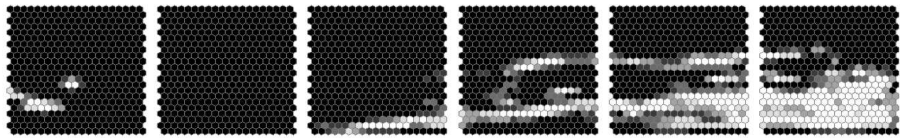}
      \caption{\label{fig:cloud_mask} \slshape A mask of grayscale values used
               in the cloud database to indicate the cloud coverage of each 
               pixel in an FD building.  Lighter values indicate greater cloud
               coverage.}
    \end{center}
  \end{figure}

  Cloud coverage at the \pao is recorded by Raytheon 2000B infrared cloud
  cameras located on the roof of each FD building.  The cameras have a spectral
  range of $7$~\micron to $14$~\micron, and photograph the field of view of the
  six FD telescopes every 5 minutes during normal data acquisition.  After the
  image data are processed, a coverage ``mask'' is created for each FD pixel,
  which can be used to remove covered pixels from the reconstruction.  Such a
  mask is shown in fig.~\ref{fig:cloud_mask}.
  
  While the IR cloud cameras record the coverage in the FD field of view, they
  cannot determine cloud heights.  The heights must be measured using the lidar
  stations, which observe clouds over each FD site during hourly
  two-dimensional scans of the atmosphere~\cite{BenZvi:2006xb}.  The Central
  Laser Facility can also observe laser echoes from clouds, though the
  measurements are more limited than the lidar observations.  Cloud height
  data from the lidar stations are combined with pixel coverage measurements to
  improve the accuracy of cloud studies.
  
  \subsection{Shoot-the-Shower}\label{subsec:sts}

\begin{figure}[ht]
    \begin{center}
      \includegraphics*[width=0.7\textwidth,clip]{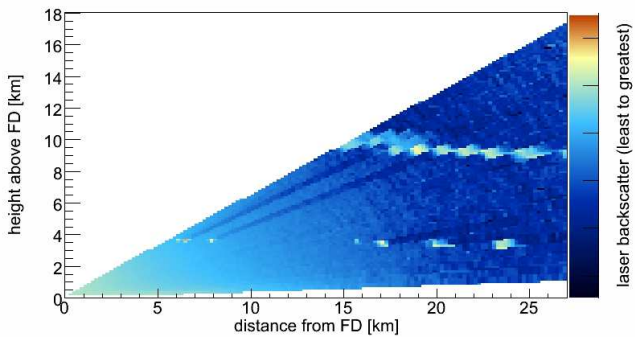}
      \caption{\label{fig:fd_sts} \slshape Lidar sweep of the shower-detector
               plane for the cloud-obscured event shown in
               fig.~\ref{fig:profile_with_cloud}.  The regions of high
               backscatter are laser echoes due to optically thick clouds.}
    \end{center}
  \end{figure}

  When a distant, high-energy air shower is detected by an FD telescope, the
  lidars interrupt their hourly sweeps and scan the plane formed by the image
  of the shower on the FD camera.  This is known as the ``shoot-the-shower''
  mode.  The shoot-the-shower mode allows the lidar station to probe for local
  atmospheric non-uniformities, such as clouds, which may affect light
  transmission between the shower and detector.  Figure~\ref{fig:fd_sts}
  depicts one of the four shoot-the-shower scans for the cloud-obscured event
  shown in fig.~\ref{fig:profile_with_cloud}.

  A preliminary implementation of shoot-the-shower was described
  in~\cite{BenZvi:2006xb}.  This scheme has been altered recently to use a fast
  on-line hybrid reconstruction now operating at the Observatory.  The new
  scheme allows for more accurate selection of showers of interest.  In
  addition, the reconstruction output can be used to trigger other atmospheric
  monitors and services, such as radiosonde balloon launches, to provide
  measurements of molecular conditions shortly after very high energy air
  showers are recorded.  ``Balloon-the-shower'' radiosonde measurements began
  at the Observatory in early 2009~\cite{Keilhauer:2009}.

%% file: acknowledgments.tex
The successful installation and commissioning of the Pierre Auger Observatory
would not have been possible without the strong commitment and effort
from the technical and administrative staff in Malarg\"ue.

We are very grateful to the following agencies and organizations for financial support: 
Comisi\'on Nacional de Energ\'ia At\'omica, 
Fundaci\'on Antorchas,
Gobierno De La Provincia de Mendoza, 
Municipalidad de Malarg\"ue,
NDM Holdings and Valle Las Le\~nas, in gratitude for their continuing
cooperation over land access, Argentina; 
the Australian Research Council;
Conselho Nacional de Desenvolvimento Cient\'ifico e Tecnol\'ogico (CNPq),
Financiadora de Estudos e Projetos (FINEP),
Funda\c{c}\~ao de Amparo \`a Pesquisa do Estado de Rio de Janeiro (FAPERJ),
Funda\c{c}\~ao de Amparo \`a Pesquisa do Estado de S\~ao Paulo (FAPESP),
Minist\'erio de Ci\^{e}ncia e Tecnologia (MCT), Brazil;
AVCR AV0Z10100502 and AV0Z10100522,
GAAV KJB300100801 and KJB100100904,
MSMT-CR LA08016, LC527, 1M06002, and MSM0021620859, Czech Republic;
Centre de Calcul IN2P3/CNRS, 
Centre National de la Recherche Scientifique (CNRS),
Conseil R\'egional Ile-de-France,
D\'epartement  Physique Nucl\'eaire et Corpusculaire (PNC-IN2P3/CNRS),
D\'epartement Sciences de l'Univers (SDU-INSU/CNRS), France;
Bundesministerium f\"ur Bildung und Forschung (BMBF),
Deutsche Forschungsgemeinschaft (DFG),
Finanzministerium Baden-W\"urttemberg,
Helmholtz-Gemeinschaft Deutscher Forschungszentren (HGF),
Ministerium f\"ur Wissenschaft und Forschung, Nordrhein-Westfalen,
Ministerium f\"ur Wissenschaft, Forschung und Kunst, Baden-W\"urttemberg, Germany; 
Istituto Nazionale di Fisica Nucleare (INFN),
Ministero dell'Istruzione, dell'Universit\`a e della Ricerca (MIUR), Italy;
Consejo Nacional de Ciencia y Tecnolog\'ia (CONACYT), Mexico;
Ministerie van Onderwijs, Cultuur en Wetenschap,
Nederlandse Organisatie voor Wetenschappelijk Onderzoek (NWO),
Stichting voor Fundamenteel Onderzoek der Materie (FOM), Netherlands;
Ministry of Science and Higher Education,
Grant Nos. 1 P03 D 014 30, N202 090 31/0623, and PAP/218/2006, Poland;
Funda\c{c}\~ao para a Ci\^{e}ncia e a Tecnologia, Portugal;
Ministry for Higher Education, Science, and Technology,
Slovenian Research Agency, Slovenia;
Comunidad de Madrid, 
Consejer\'ia de Educaci\'on de la Comunidad de Castilla La Mancha, 
FEDER funds, 
Ministerio de Ciencia e Innovaci\'on,
Xunta de Galicia, Spain;
Science and Technology Facilities Council, United Kingdom;
Department of Energy, Contract No. DE-AC02-07CH11359,
National Science Foundation, Grant No. 0450696,
The Grainger Foundation USA; 
ALFA-EC / HELEN,
European Union 6th Framework Program,
Grant No. MEIF-CT-2005-025057, 
European Union 7th Framework Program, Grant No. PIEF-GA-2008-220240,
and UNESCO.